\documentclass[11pt,a4paper]{article}
\pdfoutput=1
\usepackage{jheppub}

\usepackage{multirow, graphicx,amssymb,url,mathrsfs,amsmath}
\usepackage{wrapfig,boxedminipage,setspace,epsfig}
\usepackage{subcaption}
\usepackage{amsxtra,amstext,latexsym,dsfont,amsfonts}
\usepackage{color,eucal}
\usepackage[dvipsnames]{xcolor}
\usepackage{float}
\usepackage{slashed,comment}
\usepackage{kotex}
\usepackage{tikz}
\usetikzlibrary{calc,patterns,angles,quotes}
\usetikzlibrary{decorations.pathreplacing,decorations.markings,snakes}
\usepackage{tabularx, array}
\usepackage{mdframed,mathtools}
\newcolumntype{L}[1]{>{\raggedright\arraybackslash}p{#1}}
\newcolumntype{C}[1]{>{\centering\arraybackslash}p{#1}}
\newcolumntype{R}[1]{>{\raggedleft\arraybackslash}p{#1}}
\newcommand{\lrt}[1]{\left[ #1 \right]}

\newcommand{\nn}{\nonumber}

\newcommand{\bra}[1]{\mbox{$\langle #1 |$}}
\newcommand{\ket}[1]{\mbox{$| #1 \rangle$}}

\newcommand{\Tr}{{\rm Tr}\,}

\newcommand{\be}{\begin{equation}}
\newcommand{\ee}{\end{equation}}
\newcommand{\bea}{\begin{eqnarray}}
\newcommand{\eea}{\end{eqnarray}}

\newcommand{\lr}[1]{\left( #1 \right)}
\newcommand{\ltb}[1]{\Bigg[ #1 \Bigg]}

\title{Holographic Local Operator Quenches with Conserved Momentum and Spin}

\author[a,b,c]{Pawel Caputa,}
\author[a]{Pedro Castellini Grand,}
\author[d]{Justin R. David,}
\author[e]{Rahul Metya}
\preprint{YITP-26-110}

\affiliation[a]{
The Oskar Klein Centre and Department of Physics, Stockholm University, AlbaNova, 106 91 Stockholm, Sweden}
\affiliation[b]{Faculty of Physics, University of Warsaw, Pasteura 5, 02-093 Warsaw, Poland }
\affiliation[c]{Yukawa Institute for Theoretical Physics, Kyoto University, Kitashirakawa Oiwakecho, Sakyo-ku, Kyoto 606-8502, Japan}

\affiliation[d]{Centre for High Energy Physics, Indian Institute of Science, C. V. Raman Avenue, Bangalore
560012, India}
\affiliation[e]{International Centre for Theoretical Sciences-TIFR, Shivakote, Heseraghatta Hobli, Bangalore North 560089, India}

\abstract{We investigate the holographic dictionary relating point particles carrying longitudinal momentum or angular momentum in asymptotically AdS$_3$ spacetimes to suitably regulated, time-evolved states created by local primary operators in the dual two-dimensional conformal field theory (2D CFT). We find that asymmetric left/right Euclidean smearing of local operators produces states carrying momentum, and the corresponding bulk excitation is a particle with conserved momentum. We compute the energy density and entanglement entropy in these states and in their dual back-reacted geometries, finding exact agreement between the CFT and gravity descriptions. We further extend this correspondence to particles with intrinsic spin, whose CFT duals are primary operators with unequal holomorphic and anti-holomorphic scaling dimensions. We again find a precise match between CFT and holographic calculations of energy densities and entanglement entropies. Finally, we explore applications of these setups beyond holography by deriving the evolution of Rényi entropies in 2D rational CFTs and introducing a new class of local quantum quench protocols with conserved longitudinal or angular momentum.}
\begin{document}

\maketitle
%%%%%%%%%%%%%%%%%%%%%%%%%%%%%%%%%%%%%%%%%%%
%%%%%%%%%%%%%%%%%%%%%%%%%%%%%%%%%%%%%%%%%%%
\section{Introduction}
%%%%%%%%%%%%%%%%%%%%%%%%%%%%%%%%%%%%%%%%%%%
%%%%%%%%%%%%%%%%%%%%%%%%%%%%%%%%%%%%%%%%%%%
\,\,\,\,\,\,
 Understanding the non-equilibrium dynamics of interacting quantum many-body systems remains one of the central challenges in theoretical physics \cite{Eisert:2014jea}. Fundamental questions such as how local excitations propagate, how quantum information spreads, and under what conditions isolated quantum systems thermalize continue to attract attention across condensed matter physics, quantum many-body and quantum field theory (QFT) \cite{d2016quantum}, and quantum gravity \cite{Balasubramanian:2010ce}. While generic interacting systems are expected to lose memory of their initial conditions and approach thermal equilibrium, this paradigm is non-trivially modified by the presence of conserved quantities. Symmetries and conservation laws constrain transport, modify the propagation of correlations, and may even prevent thermalization \cite{deutsch1991quantum,Srednicki:1994mfb,DAlessio:2015qtq}, leading instead to generalized equilibrium states \cite{rigol2007relaxation}. Understanding the interplay between conserved charges, information transport, and entanglement has therefore become a key ingredient in developing a general theory of quantum dynamics far from equilibrium.

Unfortunately, these problems are notoriously difficult to study analytically. Exact real-time results are typically restricted, either to free theories or integrable models (where dynamics can be explained by a quasi-particle picture \cite{alba2017entanglement}), or systems possessing an exceptionally large degree of symmetry. Among the latter, two-dimensional conformal field theories (2D CFTs) are special. Their infinite-dimensional conformal symmetry makes it possible to compute a wide range of dynamical observables exactly, including correlation functions and entanglement measures \cite{Calabrese:2007mtj}. Consequently, 2D CFTs have become a great theoretical laboratory for exploring universal aspects of quantum quenches, transport phenomena and the propagation of excitations \cite{Bernard:2016nci,Calabrese:2016xau}.

A complementary perspective is provided by the AdS/CFT correspondence \cite{Maldacena:1997re}, which relates strongly-coupled CFTs to classical gravitational dynamics in asymptotically Anti-de Sitter (AdS) spacetimes. Within this framework, out-of-equilibrium processes acquire a remarkably simple geometric interpretation. For instance, local operator excitations in the CFT are described by massive particles propagating in AdS \cite{Nozaki:2013wia}, while quantum information measures such as entanglement entropy are mapped to geometric quantities through the Ryu--Takayanagi (RT) \cite{Ryu:2006bv} and Hubeny--Rangamani--Takayanagi (HRT) \cite{Hubeny:2007xt} prescriptions. This dual description has led to significant progress in understanding thermalization, entanglement growth, scrambling, and information propagation in strongly coupled quantum systems, while simultaneously providing valuable insights into the emergence of spacetime from quantum degrees of freedom \cite{VanRaamsdonk:2010pw,Takayanagi:2025ula}.

One of the most successful applications of this framework has been the study of local operator quenches. In holography, these processes are described by massive particles released near the AdS boundary and falling into the bulk, whose gravitational back-reaction encodes the dynamics of the excited quantum state \cite{Nozaki:2013wia,Caputa:2014vaa}. This correspondence has enabled analytic computations of time-dependent stress tensors, entanglement entropy, and other observables, establishing one of the clearest examples of real-time holography \cite{Asplund:2011cq,Nozaki:2013wia,Asplund:2013zba,Caputa:2014vaa,Asplund:2014coa,Caputa:2015tua,david2016universal,Kusuki:2017upd,Kusuki:2018wpa,Bhattacharyya:2019ifi,Mao:2024cnm,Doi:2026uqx,Mao:2025cfl,Mao:2025hkp,Bai:2026avl,Kudler-Flam:2023ahk}. Nevertheless, despite the extensive development of this subject, most existing studies have focused on excitations carrying energy but no additional conserved charges. Exceptions include \cite{Caputa:2013eka,Caputa:2015qbk,David:2017eno,Berenstein:2019tcs,David:2019bmi,David:2026owc}, which explored quenches involving conserved charges, primarily from either the bulk or the boundary perspective. 

From both the gravitational and field-theoretical perspectives, it is natural to ask how the dynamics changes when the excitation also carries momentum or angular momentum. Conserved momentum or spin is an ubiquitous ingredient of non-equilibrium phenomena, governing energy transport, ballistic propagation, and the redistribution of quantum information. On the gravity side, one natural way of incorporating such conserved charges is conceptually straightforward: one considers massive particles following geodesics with non-vanishing longitudinal velocity, angular momentum, or intrinsic spin in asymptotically AdS spacetimes.  In spite of the importance of 
studying quenches with conserved momenta or spin, it was only recently quenches carrying  non-zero longitudinal momentum were found
\cite{David:2026owc} in the context of strings infalling into the BTZ black hole. The precise and general dual description of these configurations, both in the bulk and in the boundary CFT, has not been studied in detail. 

The primary goal of this work is to develop and explore this dictionary. In the first part of this work, we consider massive particles carrying longitudinal momentum or angular momentum in asymptotically AdS$_3$ spacetimes and construct their fully back-reacted geometries analytically using embedding-space techniques and suitable two-boost transformations. We study three complementary settings: Poincaré AdS, global AdS and the BTZ black brane, which together describe vacuum, finite-size and finite-temperature states of the dual CFT. From these exact geometries we derive holographic stress tensors and time-dependent entanglement entropies in closed analytic form.

On the CFT side, we propose a universal family of locally excited states obtained by introducing independent chiral and anti-chiral regulators for local primary operators. Physically, these states describe local operator quenches carrying finite longitudinal momentum, thereby extending the standard framework of local quenches. We demonstrate that, after an explicit identification of the bulk and boundary parameters, these states exactly reproduce the holographic stress tensors in all backgrounds considered. We further perform independent computations of entanglement entropy in large-$c$ CFTs and find complete agreement with the holographic HRT prescription. 

Moreover, our discussion naturally extends from scalar to spinning excitations. After understanding how the longitudinal momentum modifies the propagation of the excitation, we turn to scenarios with intrinsic spin, which is a genuine internal quantum number. In the dual CFT this is encoded by allowing the local operator to possess unequal holomorphic and anti-holomorphic scaling dimensions, $h\neq \bar{h}$. In the bulk, the corresponding particle is described by the worldline action supplemented by the spin term, leading to a simple extension of the holographic dictionary. We again compute and match energy density and entanglement entropies in AdS and CFT for this setting.

An important aspect of our discussion is that it naturally extends beyond holography. In particular, for our first case, once the dual CFT states are identified, they define a new family of analytically tractable non-equilibrium states in arbitrary 2D CFTs, irrespective of whether a classical gravitational dual exists. To illustrate this, we investigate the dynamics of Rényi entropies in rational CFTs (RCFTs) and discuss a new class of local quantum quench protocols in which the initial excitation carries conserved momentum. These provide simple yet non-trivial settings for studying the interplay between energy transport, conserved charges, and quantum entanglement beyond the semiclassical gravity regime.

This paper is organized as follows. In Section~\ref{sec:GravitySetups}, starting with holography, we first discuss massive point particles with conserved charges in asymptotically $AdS_3$ spacetimes, construct their back-reacted metrics and compute holographic stress tensors. In Section~\ref{sec:CFTDescription}, we construct dual states in 2D CFT and compute holographic stress tensors which match the holographic results. In Section~\ref{sec:EntanglementEntropies}, we further add to the holographic dictionary by computing entanglement entropy in gravity and matching it with large-$c$ 2D CFTs. In Section~\ref{sec:BeyondHolography} we perform computations in 2D RCFTs and discuss applications of our results beyond holography. Finally, Section~\ref{sec:Conclusions} contains conclusions, and some technical details are included in the appendix.

%%%%%%%%%%%%%%%%%%%%%%%%%%%%%%%%%%%%%%%%%%%
%%%%%%%%%%%%%%%%%%%%%%%%%%%%%%%%%%%%%%%%%%%
\section{Gravity setup}\label{sec:GravitySetups}
%%%%%%%%%%%%%%%%%%%%%%%%%%%%%%%%%%%%%%%%%%%
%%%%%%%%%%%%%%%%%%%%%%%%%%%%%%%%%%%%%%%%%%%
\,\,\,\,\,\,
We begin by introducing a gravitational setup involving point particles with conserved charges propagating in AdS spacetimes. This provides the geometric setting and motivation for the analysis that follows.

Consider asymptotically $AdS_3$ geometries with radius $R$, specified by embedding coordinates
\bea
-X^2_0-X^2_1+X^2_2+X^2_3=-R^2\,,\label{eq:embedding}
\eea
and metric
\be
ds^2=-dX^2_0-dX^2_1+dX^2_2+dX^2_3\,.
\ee
Below, we will be interested in three particular coordinates: the Poincaré, the Global, and BTZ black brane (specified below), given by $x^\mu$, $\mu=\{0,1,2\}$, with real time, radial direction, and one spatial/angular coordinate. After parametrizing the embedding coordinates $X_i(x^\mu)$ we will get the induced metric for these three cases in the form
\be
ds^2=g_{\mu\nu}(x)dx^\mu dx^\nu\,.
\ee
In these geometries, we will study massive point particles following timelike trajectories $x^\mu(\lambda)$, and described by world-line actions
\be
S_m=-m\int ds=-m\int \sqrt{-g_{\mu\nu}\dot{x}^\mu(\lambda) \dot{x}^\nu(\lambda)}d\lambda\,,\label{SmAction}
\ee
where $m$ is the particle's mass and parameter $\lambda$ can be chosen conveniently as time, or a more general (affine) parameter.

In general, our particles will start from some initial radial position close to the asymptotic boundary and will have a vanishing initial radial velocity and non-trivial spatial/angular momentum. As time progresses, they will be falling radially into the interior of the AdS geometry but their path in the spatial direction will also be non-trivial. Hence, they will have conserved energy as well as conserved longitudinal/angular momentum, and their trajectories will be solutions of the geodesic equations derived by extremizing \eqref{SmAction}. Our main goal will be to compute the back-reaction from the particles on the $AdS_3$ backgrounds. This amounts to solving Einstein's equations
\be
R_{\mu\nu}-\frac{1}{2}R g_{\mu\nu}+\Lambda g_{\mu\nu}=T^m_{\mu\nu}\,,
\ee
where $\Lambda_d=-d(d-1)/(2R^2)$ (in our case, $d=2$) is the cosmological constant and $T^m_{\mu\nu}$ is the stress tensor from the massive particle's action \eqref{SmAction}.

It turns out that we can construct the back-reacted geometry analytically using the following trick \cite{Horowitz:1999gf,Nozaki:2013wia}.
First, note that the SO$(2,2)$ isometry of the embedding space \eqref{eq:embedding} allows us to introduce an equivalent parametrization of a given metric with two extra boost parameters $\eta_1$ and $\eta_2$
\bea
\tilde{X}_0&=&X_0\cosh(\eta_1)-X_3\sinh(\eta_1)\,,\nn\\
\tilde{X}_1&=&X_1\cosh(\eta_2)-X_2\sinh(\eta_2)\,,\nn\\
\tilde{X}_2&=&X_2\cosh(\eta_2)-X_1\sinh(\eta_2)\,,\nn\\
\tilde{X}_3&=&X_3\cosh(\eta_1)-X_0\sinh(\eta_1)\,.\label{MapGen}
\eea 
We can use it to map the particle in a given, original $AdS_3$ metric in which we want to construct the back-reaction (embedding coordinates $X_i$), to a configuration where the massive particle is at rest at the origin of global $AdS_3$ coordinates (embedding coordinates $\tilde{X}_i$). In doing so, we fix the relation between the boost parameters and the physical parameters of the original setup.

Then, we can first compute the back-reacted metric from  the massive particle at the origin of global AdS which reads
\be
ds^2=-(r^2+R^2-M)d\tau^2+\frac{R^2dr^2}{r^2+R^2-M}+r^2d\theta^2\,.\label{BRMetric}
\ee
It describes conical singularities with $M<R^2$ or a BTZ black string for $M>R^2$. The parameter $M$ is related to the mass of the particle by
\be
m=\frac{M}{8G_NR^2}\,,\label{Mvsm}
\ee
where $G_N$ is the Newton constant.

Finally, we can again apply the map \eqref{MapGen}, but now to the back-reacted metric \eqref{BRMetric}. This will yield the fully back-reacted, asymptotically AdS spacetime from the massive particle in our original coordinates. From such constructed geometries, we will be able to extract holographic stress tensors and match them, by fixing the boost parameters or the particle's conserved charges, with dual CFT states and CFT computations of the stress tensor expectation values. Moreover, using the HRT \cite{Hubeny:2007xt} prescription, we will be able to compute holographic entanglement entropies in the back-reacted geometries and match them with large-$c$ CFT results.

Let us stress again that this procedure has already been studied extensively in the context of holographic quantum quenches, starting from the pioneering works of \cite{Nozaki:2013wia}. Here we generalize it to particles with conserved momenta. In the following subsections, we will first describe the gravity computations for the three above-mentioned examples in detail and derive the holographic stress tensor in each case. Then, in section \ref{sec:CFTDescription} we will reproduce these computations in 2D CFTs.
%%%%%%%%%%%%%%%%%%%%%%%%%%%%%%%%%%%%%%%%%%%
\subsection{Poincaré AdS}
%%%%%%%%%%%%%%%%%%%%%%%%%%%%%%%%%%%%%%%%%%%
\,\,\,\,\,\,
We start by considering the following embedding 
\bea
\sqrt{r^2+R^2}\cos(\tau)&=&\tilde{X}_0=\frac{e^{-\eta_1}(z^2+x^2-t^2)+R^2e^{\eta_1}}{2z}\,,\nn\\
\sqrt{r^2+R^2}\sin(\tau)&=&\tilde{X}_1=R\frac{t\cosh(\eta_2)-x \sinh(\eta_2)}{z}\,,\nn\\
r\sin(\theta)&=&\tilde{X}_2=R\frac{x\cosh(\eta_2)-t \sinh(\eta_2)}{z}\,,\nn\\
r\cos(\theta)&=&\tilde{X}_3=\frac{e^{-\eta_1}(z^2+x^2-t^2)-R^2e^{\eta_1}}{2z}\,,\label{MapPoincare}
\eea
which gives the induced metric on $AdS_3$ in the Poincaré form
\be
ds^2=R^2\frac{-dt^2+dx^2+dz^2}{z^2}\,.\label{MetricP}
\ee
The global metric (left embedding), in which the particle will be at rest at the origin $r=0$, is given by
\be
ds^2=-(r^2+R^2)d\tau^2+\frac{R^2dr^2}{r^2+R^2}+r^2d\theta^2\,.\label{eq:MetGlobG}
\ee
%%%%%%%%%%%%%%%%%%%%%%%%%%%%%%%%%%%%%%%%%%%
\subsubsection{Point particle with longitudinal velocity}
%%%%%%%%%%%%%%%%%%%%%%%%%%%%%%%%%%%%%%%%%%%
\,\,\,\,\,\,
First, we want to use the above relations \eqref{MapPoincare} to map a trajectory of a massive particle with energy $E$ and spatial momentum $P$ in Poincaré coordinates into a particle at $r=0$ in global coordinates \eqref{eq:MetGlobG}. Therefore, after setting to zero the last two relations in \eqref{MapPoincare}, we get two constraints
\be
x(t)=t\tanh(\eta_2)\,,\qquad z(t)=\sqrt{t^2\cosh^{-2}(\eta_2)+e^{2\eta_1}R^2}\,.\label{eq:ConstrintsPoincare}
\ee
Moreover, the relation between the time $\tau$ in global coordinates and the Poincaré time $t$ is
\be
\tan(\tau)=\frac{t}{Re^{\eta_1}\cosh(\eta_2)}\,,
\ee
while $\theta$ is mapped to $0$. We can verify that these constraints parametrize a geodesic $(x(t),z(t))$ of a massive particle in metric \eqref{MetricP}. Indeed, functions \eqref{eq:ConstrintsPoincare} solve the equations of motion derived from the action
\be
S_m=-mR\int dt\frac{\sqrt{1-x'(t)^2-z'(t)^2}}{z(t)}\equiv\int dt\mathcal{L}\,.
\ee
Now, we should fix the boost parameters carefully. Observe that from the above action we can derive the canonical momentum
\be
P\equiv p_x=\frac{\partial \mathcal{L}}{\partial x'(t)}=\frac{mRx'(t)}{z(t)\sqrt{1-x'(t)^2-z'(t)^2}}= m e^{-\eta_1}\sinh(\eta_2)\,,
\ee
where in the last step we inserted the on-shell solutions \eqref{eq:ConstrintsPoincare}. We chose the sign of $P$ to be positive and the particle moving to the right from the insertion point.

More generally, if we parametrize the timelike geodesic by $(t(s),x(s),z(s))$, it extremizes the action
\be
S_m=-mR\int ds\frac{\sqrt{t'(s)^2-x'(s)^2-z'(s)^2}}{z(s)}\,,
\ee
from which the canonical momenta 
\be
p_\mu=\frac{mg_{\mu\nu}\dot{x}^\nu}{\sqrt{-g_{\mu\nu}\dot{x}^\mu\dot{x}^\nu}}\,,
\ee
satisfy the mass-shell constraint
\be
g^{\mu\nu}p_\mu p_\nu+m^2=0\,.\label{ConstPart}
\ee
After defining the conserved energy and momentum as usual
\be
E=-p_t\,,\qquad P=p_x\,,\qquad E^2-P^2>0\,,
\ee
the constraint \eqref{ConstPart} becomes
\be
z^2(-E^2+P^2+p^2_z)+m^2R^2=0\,.\label{eq:massShellP}
\ee
The above relations are sufficient to fix the boost parameters in terms of the physical data. More precisely, the initial conditions for the trajectory of our massive particle falling from the boundary of AdS are\footnote{in this parametrization we automatically have
\be
z'(0)=0\,.
\ee} 
\be
x(t=0)=0,\qquad x'(0)=v,\qquad  z(t=0)=\epsilon,\qquad z'(0)=0\,.\label{BCPoincare}
\ee
From $z(0)=\epsilon$ and $x'(0)=v$, we fix
\be
Re^{\eta_1}=\epsilon,\qquad \tanh(\eta_2)=v\,.\label{eq:constr1}
\ee
On the other hand, at the release point of the particle, for some $s=s_*$, we have $z(s_*)=\epsilon$ and $p_z(s_*)\sim z'(s_*)=0$, so from the mass-shell condition \eqref{eq:massShellP} we find 
\be
\sqrt{E^2-P^2}=\frac{mR}{\epsilon}=me^{-\eta_1}\,,
\ee
where in the second step we inserted the first equation from \eqref{eq:constr1}.
We can solve these constraints to find\footnote{From the above formulas we also find the relation between boost parameters and conserved charges $\tanh(\eta_2)=P/E$ and $e^{-\eta_1}=\frac{\sqrt{E^2-P^2}}{m}$. }
\be
E=me^{-\eta_1}\cosh(\eta_2)=\frac{mR}{\epsilon\sqrt{1-v^2}}\,,\qquad P=me^{-\eta_1}\sinh(\eta_2)=\frac{mRv}{\epsilon\sqrt{1-v^2}}\,,\label{EPboosts}
\ee
giving the relativistic relation $v=P/E$.

\begin{figure}[h!]
    \centering
    \includegraphics[width=\linewidth]{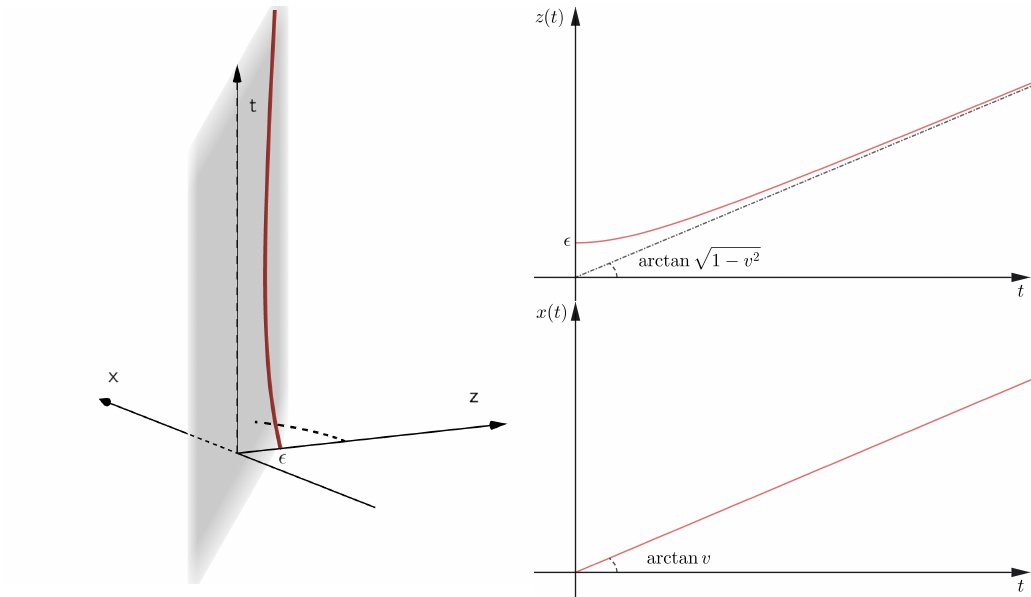}
    \caption{Particle's trajectory in Poincaré coordinates. The left panel shows the worldline (red) parametrised as $(z(t),x(t),t)$, originating at the cut-off surface $z=\epsilon$. The gray plane is marked to show that, at late times, the particle effectively moves in this slice of $AdS_3$. The right panels display the time dependence of the coordinates $z(t)$ and $x(t)$, with (asymptotic) slopes determined by $\sqrt{1-v^2}$ and ${v}$, respectively.}
    \label{fig:PointParticlePoincare}
\end{figure}

Finally, we have the particle's trajectory in terms of its charges and the cut-off
\be
x(t)=vt=\frac{P}{E}t\,,\qquad z(t)=\sqrt{\left(1-v^2\right)t^2+\epsilon^2}=\sqrt{\left(1-\frac{P^2}{E^2}\right)t^2+\epsilon^2}\,.\label{eq:TrajectoryP}
\ee
We plot this trajectory on Fig.\,\ref{fig:PointParticlePoincare}. Once we have fixed these relations, we can follow the maps and compute the back-reacted metric in Poincaré coordinates. In general, the back-reacted metric has a complicated form of a shock-wave (see e.g. \cite{Horowitz:1999gf} and references therein) supported on the trajectory of the particle \eqref{eq:TrajectoryP}\footnote{Note that our back-reacted solution is valid for arbitrary $\epsilon$ and the standard shock-wave metrics correspond to $\epsilon \to 0$. See more in  \cite{Caputa:2015waa,Afkhami-Jeddi:2017rmx}.}. For our purposes, we will not need the explicit form of these solutions in Poincaré coordinates, and we will extract the physical quantities from \eqref{BRMetric} as well as \eqref{MapPoincare} with relations \eqref{EPboosts}. In the following subsection, we start with computing the holographic stress tensor which amounts to writing the back-reacted metric in the Fefferman-Graham (FG) form \cite{Balasubramanian:1999re,deHaro:2000vlm}.
%%%%%%%%%%%%%%%%%%%%%%%%%%%%%%%%%%%%%%%%%%%
\subsubsection{Holographic stress tensor}
%%%%%%%%%%%%%%%%%%%%%%%%%%%%%%%%%%%%%%%%%%%
\,\,\,\,\,\,
To derive the holographic stress tensor we simply apply the map \eqref{MapPoincare} to the back-reacted metric in global coordinates \eqref{BRMetric}. In addition, we will follow a simpler procedure described in \cite{David:2026owc}. Namely, to extract the time dependent expectation value of the stress tensor, it is sufficient to work to the first order in $M$ in \eqref{BRMetric} and write it as
\be
ds^2=-(r^2+R^2)d\tau^2+\frac{R^2dr^2}{r^2+R^2}+r^2d\theta^2+M\left(d\tau^2+\frac{R^2dr^2}{(r^2+R^2)^2}\right)+O(M^2)\,.
\ee
Then, we apply the map \eqref{MapPoincare} and expand the resulting metric to the second order in $z\sim0$. The back-reacted metric obtained in this way is not yet in the FG form, so we must first redefine the radial coordinate as
\be
z\to z(1-\alpha z^2)\,,
\ee
and fix $\alpha$ such that, to the leading order in $z\sim 0$, the metric becomes
\be
ds^2\sim R^2\frac{dz^2+g_{ij}(x,z)dx^idx^j}{z^2}\,,\qquad g_{ij}(x,z)=g^{(0)}_{ij}+g^{(2)}_{ij}z^2\,.
\ee
In this way, from the sub-leading correction $g^{(2)}_{ij}$, we can read off the holographic stress tensor as
\be
T_{ij}=\frac{R}{8\pi G_N}g^{(2)}_{ij}\,.
\ee
This procedure is in fact general and will be applied below to all the three cases. What changes in specific examples is the dependence of $\alpha$ on the physical parameters and boundary coordinates (by definition it does not depend on the radial coordinate).

\begin{figure}[h!]
	\vspace{0.3cm}
	\begin{center}
    \includegraphics[height=0.26\linewidth]{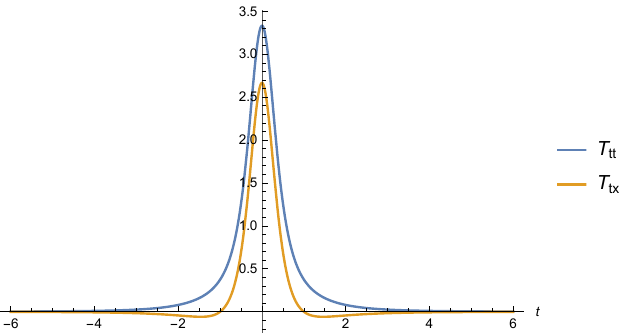}\hspace{1mm}
		\includegraphics[height=0.26\linewidth]{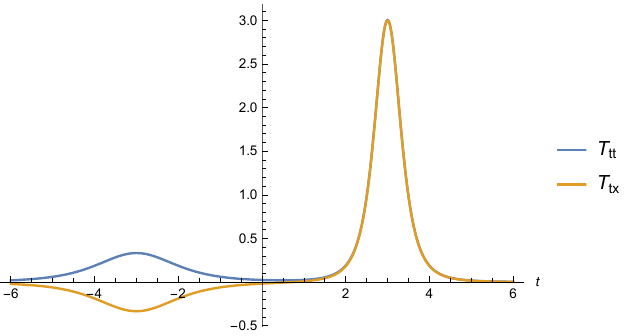}
	\end{center}
	\caption{Time evolution of energy and momentum densities (normalized by $mR/\pi$) for $t=0$ (left) and $t=3$ (right). Plot for $\epsilon=1$ and $\tanh(\eta_2)=1/2$. On the right plot, the profiles of $T_{tt}$ and $T_{tx}$ moving to the right coincide, up to our numerical precision.}	\label{fig:EnMomPoincare}
\end{figure}

More precisely, in the Poincaré coordinates, we find the key step in the redefinition of the z-coordinate
\be
z\to z\left(1-\frac{M^2R^2 e^{2\eta_1}}{(x^2_-+R^2e^{2(\eta_1-\eta_2)})(x^2_++R^2e^{2(\eta_1+\eta_2)})}z^2\right)\,,\label{zalphaP}
\ee
where we introduced light cone coordinates $x_\pm=x\pm t$. In this way, employing the map \eqref{MapPoincare} and coordinate \eqref{zalphaP}, we find that the only non-vanishing components of the holographic stress tensor are along the light cone directions
\be
T_{--}(x_-)=\frac{MR}{8\pi G_N}\frac{e^{2(\eta_1-\eta_2)}}{(x^2_-+R^2e^{2(\eta_1-\eta_2)})^2}\,,\qquad T_{++}(x_+)=\frac{MR}{8\pi G_N}\frac{e^{2(\eta_1+\eta_2)}}{(x^2_++R^2e^{2(\eta_1+\eta_2)})^2}\,.\label{HSTPoincare}
\ee
We plot the energy and momentum densities on Fig.\,\ref{fig:EnMomPoincare}. Initially, we can see the two pulses of different heights determined by the boost parameters\footnote{given by $\frac{2mR}{\pi\epsilon^2}\cosh(2\eta_2)$ for the energy and $\frac{2mR}{\pi\epsilon^2}\sinh(2\eta_2)$ for the momentum densities.} at $t=0$ which then split and move to the left and to the right. The peaks for the energy density are positive but, from the explicit expressions, we clearly see that the left moving momentum density is negative (mirror image of the left moving energy peak).

Finally, using \eqref{Mvsm}, we can compute the total energy and momentum of the gravity solution by integrating $T_{tt}(x)$ and $T_{tx}(x)$ on some fixed time-slice as follows
\be
E=\int dx\, T_{tt}(x)=\int dx (T_{++}(x)+T_{--}(x))=\frac{mR}{\epsilon}\cosh(\eta_2)\,,\label{TEnP}
\ee
\be
P=\int dx\, T_{tx}(x)=\int dx (T_{--}(x)-T_{++}(x))=\frac{mR}{\epsilon}\sinh(\eta_2)\,.\label{TPP}
\ee
These are of course consistent with \eqref{EPboosts}.
%%%%%%%%%%%%%%%%%%%%%%%%%%%%%%%%%%%%%%%%%%%
\subsection{Global AdS}
%%%%%%%%%%%%%%%%%%%%%%%%%%%%%%%%%%%%%%%%%%%
\,\,\,\,\,\,
We can repeat the analogous analysis in global coordinates. Similarly to the Poincaré example above, we will map the point particle with angular momentum in global coordinates to an auxiliary global metric where the particle is at rest at the origin $r=0$. To avoid confusion, we use a slightly different parametrization of the starting global metric and the same auxiliary global metric as in the previous example. Namely, we consider the map \eqref{MapGen} with global coordinates 
\bea
\sqrt{r^2+R^2}\cos(\tau)&=&X_0=R\cosh(\rho)\cos\left(\frac{2\pi t}{L}\right)\,,\nn\\
\sqrt{r^2+R^2}\sin(\tau)&=&X_1=R\cosh(\rho)\sin\left(\frac{2\pi t}{L}\right)\,, \\ 
r\sin(\theta)&=&X_2=R\sinh(\rho)\sin\left(\frac{2\pi \phi}{L}\right)\,,\nn\\
r\cos(\theta)&=&X_3=R\sinh(\rho)\cos\left(\frac{2\pi \phi}{L}\right)\,,\label{MAPGLOBALC}
\eea
where the left hand side is the same as before and the coordinates on the right hand side are inserted into \eqref{MapGen}. This choice will be also more convenient for comparisons with CFT results on a cylinder of spatial circumference $L$. The global metric in these coordinates is given by
\be
ds^2=R^2\left[d\rho^2+\frac{4\pi^2}{L^2}\left(-\cosh^2(\rho)dt^2+\sinh^2(\rho)d\phi^2\right)\right]\,.
\ee
%%%%%%%%%%%%%%%%%%%%%%%%%%%%%%%%%%%%%%%%%%%
\subsubsection{Point particle with angular momentum}
%%%%%%%%%%%%%%%%%%%%%%%%%%%%%%%%%%%%%%%%%%%
\,\,\,\,\,\,
Next, we determine the geodesic and the relation between the boosts and physical parameters. To map the particle with angular momentum in global coordinates $(\rho,t,\phi)$ into the $(r=0,\theta=0)$ of global coordinates, we again employ the transformation \eqref{MapGen}. This fixes the two relations
\be
\tan\left(\frac{2\pi\phi(t)}{L}\right)=\frac{\tanh(\eta_2)}{\tanh(\eta_1)}\tan\left(\frac{2\pi t}{L}\right)\,,\qquad \tanh(\rho(t))=\frac{\sin\left(\frac{2\pi t}{L}\right)}{\sin\left(\frac{2\pi \phi(t)}{L}\right)}\tanh(\eta_2)\,,\label{eq:ParamGlobal}
\ee
as well as the map between the two times
\be
\tan(\tau)=\frac{\cosh(\eta_1)}{\cosh(\eta_2)}\tan\left(\frac{2\pi t}{L}\right)\,.
\ee
We can again verify that the trajectory $(\rho(t),\phi(t))$ given by \eqref{eq:ParamGlobal} solves the e.o.m coming from the action
\be
S_m=-mR\int dt\sqrt{\frac{4\pi^2}{L^2}(\cosh^2(\rho(t))-\sinh^2(\rho(t))\phi'(t)^2)-\rho'^2(t)}\,.
\ee
Moreover, the conserved (canonical) angular momentum is given by\footnote{We also choose it to be positive.}
\be
J\equiv p_\phi=\frac{2\pi m R}{L}\sinh(\eta_1)\sinh(\eta_2)\,.\label{Jalfa}
\ee

\begin{figure}[h!]
    \centering
    \includegraphics[width=\linewidth]{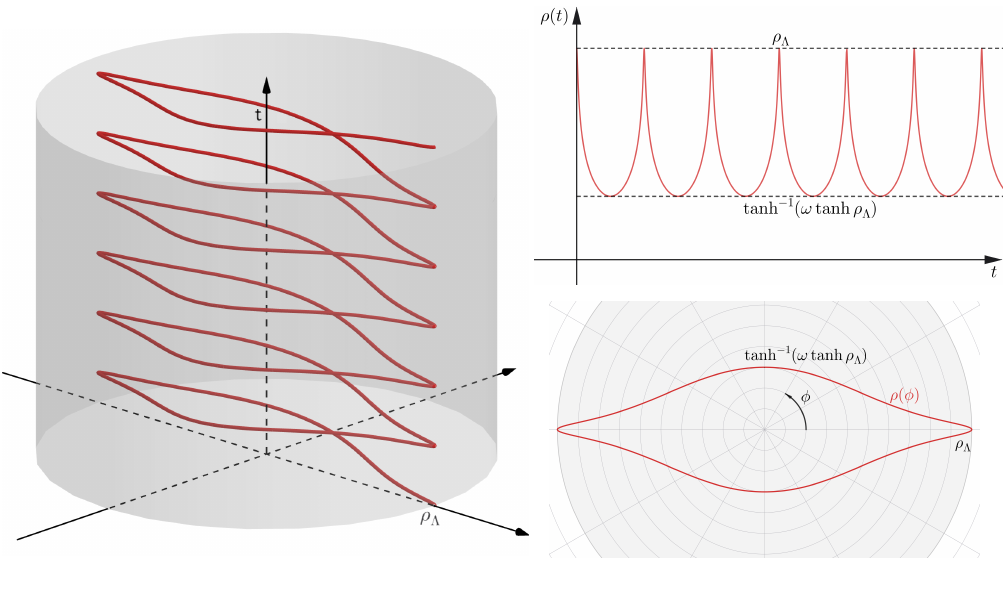}
    \caption{Particle's trajectory in Global Coordinates. The left panel shows the worldline (red) parametrised as $(\rho(t)\cos{\phi(t)},\rho(t)\sin{\phi(t)},t)$, originating at the cutoff surface $\rho=\rho_\Lambda$. The right panels display the time dependence of the coordinate $\rho$ and the trajectory given as $\rho(\phi)$, respectively}
    \label{fig:PointParticleGlobal}
\end{figure}

The most general geodesic's parametrization $(t(s),\phi(s),\rho(s))$ satisfies equations of motion from the action
\be
S_m=-mR\int ds\sqrt{\frac{4\pi^2}{L^2}(\cosh^2(\rho(s))t'(s)^2-\sinh^2(\rho(s))\phi'(s)^2)-\rho'(s)^2}\,,
\ee
and the mass-shell condition \eqref{ConstPart} for its canonical momenta becomes
\be
p^2_\rho-\frac{L^2}{4\pi^2}\left(\frac{E^2}{\cosh^2\rho}-\frac{J^2}{\sinh^2\rho}\right)=-m^2R^2\,.
\ee
This time, the boundary conditions for the particle are
\be
\phi(0)=0\,,\qquad \phi'(0)=\omega\,,\qquad \rho(0)=\rho_\Lambda\,,\qquad \rho'(0)=0\,.
\ee
It is easy to check that the first and last condition are automatically satisfied by \eqref{eq:ParamGlobal}. The initial condition for $\rho(0)$ fixes the first boost parameter
\be
\eta_1=\rho_\Lambda\,,
\ee
while the angular velocity is fixed to
\be
\omega=\frac{\tanh(\eta_2)}{\tanh(\eta_1)}\,.
\ee
It is also useful to determine the conserved energy in terms of the boost parameters. For that, we can again use the fact that at the release point near the boundary at $\rho=\rho_\Lambda$ we have $p_\rho=0$, so from the mass-shell condition we have
\be
\frac{E^2}{\cosh^2\rho_\Lambda}-\frac{J^2}{\sinh^2\rho_\Lambda}=\frac{4\pi^2m^2R^2}{L^2}\,.
\ee
Together with \eqref{Jalfa} we then find
\be
E=\frac{2\pi m R}{L}\cosh(\eta_1)\cosh(\eta_2)\,,\qquad J=\frac{2\pi m R}{L}\sinh(\eta_1)\sinh(\eta_2)\,.\label{EJGlobalp}
\ee
Finally, we can write the geodesic in terms of these parameters as
\be
\tan\left(\frac{2\pi\phi(t)}{L}\right)=\omega\tan\left(\frac{2\pi t}{L}\right)=\frac{J}{E}\frac{\tan\left(\frac{2\pi t}{L}\right)}{\tanh^2(\rho_\Lambda)}\,,
\ee
and similarly 
\bea
\tanh(\rho(t))&=&\cos\left(\frac{2\pi t}{L}\right)\tanh(\rho_\Lambda)\sqrt{1+\omega^2\tan^2\left(\frac{2\pi t}{L}\right)}\nn\\
&=&\cos\left(\frac{2\pi t}{L}\right)\tanh(\rho_\Lambda)\sqrt{1+\frac{J^2\tan^2\left(\frac{2\pi t}{L}\right)}{E^2\tanh^4(\rho_\Lambda)}}\,.
\eea
Plots for this geodesic can be found on Fig.\,\ref{fig:PointParticleGlobal}. For $\omega=J=0$ this generalizes geodesics recently studied in \cite{Caputa:2024sux}.

%%%%%%%%%%%%%%%%%%%%%%%%%%%%%%%%%%%%%%%%%%%
\subsubsection{Holographic stress tensor}
%%%%%%%%%%%%%%%%%%%%%%%%%%%%%%%%%%%%%%%%%%%
\,\,\,\,\,\,
Next we compute the holographic stress tensor in the back-reacted geometry. We follow the same procedure as for the Poincaré example and just work with the back-reacted metric in global coordinates to the first order in $M$. We then apply the map \eqref{MapGen} with global coordinates on the right of \eqref{MAPGLOBALC}. The key step is again to fix the FG coordinate which is given here by
\be
\rho=-\log(z(1-\alpha z^2))\,,
\ee
with parameter $\alpha$ expressed as
\be
\alpha=\frac{Me^{2\eta_1}}{R^2\left(\cos^2\left(\frac{\pi x_-}{L}\right)+e^{2(\eta_1+\eta_2)}\sin^2\left(\frac{\pi x_-}{L}\right)\right)\left(\cos^2\left(\frac{\pi x_+}{L}\right)+e^{2(\eta_1-\eta_2)}\sin^2\left(\frac{\pi x_+}{L}\right)\right)}\,,
\ee
where we again used the light cone coordinates on the cylinder $x_\pm=t\pm\phi$.

After these steps, we arrive at the only non-trivial components of the stress tensor along the light cone directions
\bea
T_{\pm\pm}(x_\pm)=-\frac{\pi R}{8G_N L^2}+\frac{\pi M}{8G_N L^2 R}\frac{1}{\left(\cosh(\eta_1\mp\eta_2)-\cos\left(\frac{2\pi x_{\pm}}{L}\right)\sinh(\eta_1\mp \eta_2)\right)^2}\,.\label{eqn:STGlobalHol}
\eea
This is one of the main results of this subsection and we will return to it in Section \ref{sec:CFTDescription}. 

Last but not least, we can compute the total energy and angular momentum from the holographic stress tensor in the back-reacted metric. Analogously to \eqref{TEnP} and \eqref{TPP}, we find them to be
\bea
E=-\frac{\pi R}{4G_N L}+\frac{\pi M}{4G_N L R}\cosh(\eta_1)\cosh(\eta_2)&=&-\frac{\pi R}{4G_N L}+\frac{2\pi mR}{L}\cosh(\eta_1)\cosh(\eta_2)\,,\nonumber\\
J=\frac{\pi M}{4G_N L R}\sinh(\eta_1)\sinh(\eta_2)&=&\frac{2\pi mR }{ L }\sinh(\eta_1)\sinh(\eta_2)\,,
\eea
where in the right hand side we used \eqref{Mvsm}. Clearly, these expressions reproduce \eqref{EJGlobalp} together with the Casimir energy of the global $AdS_3$ spacetime (the negative terms in the first line).  
%%%%%%%%%%%%%%%%%%%%%%%%%%%%%%%%%%%%%%%%%%%
\subsection{BTZ black brane}
%%%%%%%%%%%%%%%%%%%%%%%%%%%%%%%%%%%%%%%%%%%
\,\,\,\,\,\,
Finally, we consider the example of an infalling massive particle with longitudinal velocity propagating in the BTZ black brane background. Although this example was already analyzed in \cite{David:2026owc}, we revisit it here using our uniform conventions in order to keep the presentation self-contained.

We start with the following map from the BTZ black brane to global coordinates  
\bea
\sqrt{r^2+R^2}\cos(\tau)&=&X_0=\frac{R\beta}{2\pi z}\cosh\left(\frac{2\pi x}{\beta}\right)\,,\nn\\
\sqrt{r^2+R^2}\sin(\tau)&=&X_1=\frac{R\beta}{2\pi z}\sqrt{1-\frac{4\pi^2}{\beta^2}z^2}\sinh\left(\frac{2\pi t}{\beta}\right)\,,\nn\\
r\sin(\theta)&=&X_2=\frac{R\beta}{2\pi z}\sinh\left(\frac{2\pi x}{\beta}\right)\,,\nn\\
r\cos(\theta)&=&X_3=\frac{R\beta}{2\pi z}\sqrt{1-\frac{4\pi^2}{\beta^2}z^2}\cosh\left(\frac{2\pi t}{\beta}\right)\,.\label{MapPSBTZ}
\eea
The induced metric can be written as
\be
ds^2=\frac{R^2}{z^2}\left(-(1-\tilde{M}z^2)dt^2+\frac{dz^2}{1-\tilde{M}z^2}+dx^2\right)\,,\qquad \sqrt{\tilde{M}}=\frac{2\pi}{\beta}\,,\label{eqn:MetricBTZ}
\ee
where $\beta=1/T$ is the inverse temperature and $x$ is the non-compact spatial direction. The asymptotic boundary is at $z=0$ and the horizon at $z=\tilde{M}^{-1/2}=\beta/(2\pi)$. As before, we will employ the map with two boosts \eqref{MapGen} and find the point particle trajectory as well its back-reaction on \eqref{eqn:MetricBTZ}. This will allow us to extract the holographic stress tensor. 
%%%%%%%%%%%%%%%%%%%%%%%%%%%%%%%%%%%%%%%%%%%
\subsubsection{Point particle with longitudinal velocity}
%%%%%%%%%%%%%%%%%%%%%%%%%%%%%%%%%%%%%%%%%%%
\,\,\,\,\,\,
First, we fix the map between the infalling particle with longitudinal velocity in BTZ and the setup where the particle is at rest at the origin of global coordinates. Taking the general map \eqref{MapGen} with coordinates \eqref{MapPSBTZ},  we start by setting the last two relations to zero, so that $r=0$ in global coordinates. This yields the two constraints
\be
\tanh\left(\frac{2\pi x(t)}{\beta}\right)=\tanh\left(\frac{2\pi t}{\beta}\right)\tanh(\eta_1)\tanh(\eta_2)\,,
\ee
and
\be
z(t)=\frac{\beta}{2\pi}\sqrt{1-\frac{\cosh^2\left(\frac{2\pi x(t)}{\beta}\right)}{\cosh^2\left(\frac{2\pi t}{\beta}\right)}\tanh^2(\eta_1)}\,,
\ee
as well as the relation between times
\be
\tan(\tau)=\frac{\sinh(\eta_1)}{\cosh(\eta_2)}\tanh\left(\frac{2\pi t}{\beta}\right)\,.
\ee
Again, this trajectory parametrized by $(x(t),z(t))$ is a geodesic extremizing the action
\be
S_m=-mR\int \frac{dt}{z(t)}\sqrt{1-\tilde{M}z(t)^2-x'(t)^2-\frac{z'(t)^2}{1-\tilde{M}z(t)^2}}\,.
\ee
The conserved canonical momentum is now given in terms of the boost parameters as\footnote{As before, we fix the particle's motion to have positive $P$.}
\be
P\equiv p_x=\frac{2\pi m R}{\beta}\cosh(\eta_1)\sinh(\eta_2)\,.%=\frac{mR}{\epsilon}\sinh(\eta_2)\,.
\ee
Moreover, the mass shell condition (derived from the canonical momenta of the action with $(t(s),x(s),z(s))$) is now
\be
z^2\left(-\frac{E^2}{1-\tilde{M}z^2}+P^2+p^2_z(1-\tilde{M}z^2)\right)=-m^2R^2\,.
\ee
The boundary conditions for the in-falling particle are exactly the same as in \eqref{BCPoincare}. They provide two non-trivial constraints
\be
\cosh(\eta_1)=\frac{\beta}{2\pi\epsilon}\,,\qquad v=\tanh(\eta_1)\tanh(\eta_2)\,.
\ee

\begin{figure}[h!]
    \centering
    \includegraphics[width=\linewidth]{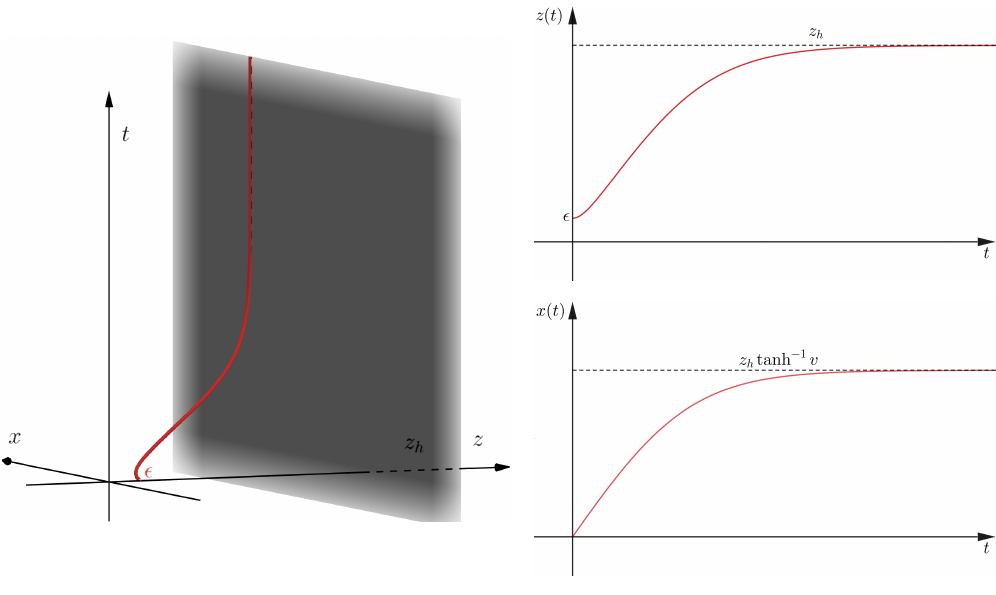}
    \caption{Particle's trajectory in the BTZ background. The left panel shows the worldline (red) parametrised as $(z(t),x(t),t)$, originating at the cut-off surface $z=\epsilon$. The black plane represents the BTZ horizon at $z=z_h$. The right panels display the time dependence of the coordinates $z(t)$ and $x(t)$, with asymptotes given by $z_h$ and $z_h\arctan{v}$, respectively.}
    \label{fig:PointParticleBTZ}
\end{figure}

To fix the energy in terms of the boosts we use that at the release point of the particle we have $p_z=0$ and $z=\epsilon$ such that the mass-shell condition reduces to
\be
\frac{E^2}{1-\tilde{M}\epsilon^2}-P^2=\frac{m^2R^2}{\epsilon^2}\,.
\ee
This fixes the two relations
\be
E=\frac{2\pi mR}{\beta}\sinh(\eta_1)\cosh(\eta_2)\,,\qquad P=\frac{2\pi m R}{\beta}\cosh(\eta_1)\sinh(\eta_2)\,.
\ee
Finally, we can write the geodesic in terms of the physical parameters as
\be
\tanh\left(\frac{2\pi x(t)}{\beta}\right)=v\tanh\left(\frac{2\pi t}{\beta}\right)=\frac{P}{E}\left(1-\frac{4\pi^2\epsilon^2}{\beta^2}\right)\tanh\left(\frac{2\pi t}{\beta}\right)\,,
\ee
and
\be
z(t)=\frac{\beta}{2\pi}\sqrt{1-\frac{\left(1-\frac{4\pi^2\epsilon^2}{\beta^2}\right)\cosh^{-2}\left(\frac{2\pi t}{\beta}\right)}{1-v^2\tanh^2\left(\frac{2\pi t}{\beta}\right)}}=\frac{\beta}{2\pi}\sqrt{1-\frac{\left(1-\frac{4\pi^2\epsilon^2}{\beta^2}\right)\cosh^{-2}\left(\frac{2\pi t}{\beta}\right)}{1-\frac{P^2}{E^2}\left(1-\frac{4\pi^2\epsilon^2}{\beta^2}\right)^2\tanh^2\left(\frac{2\pi t}{\beta}\right)}}\,.
\ee
For $P=0$ this reproduces the trajectory from \cite{Caputa:2014eta}. For $t\to\infty$ the particle approaches the horizon at $z=z_h\equiv\beta/(2\pi)$, as can be seen on Fig.\,\ref{fig:PointParticleBTZ}. 

Using the maps and our prescription we can explicitly construct the back-reacted shock-wave geometry. Below we will directly evaluate the holographic stress tensor from this solution.
%%%%%%%%%%%%%%%%%%%%%%%%%%%%%%%%%%%%%%%%%%%
\subsubsection{Holographic stress tensor}
%%%%%%%%%%%%%%%%%%%%%%%%%%%%%%%%%%%%%%%%%%%
\,\,\,\,\,\,
Following the same procedure as in the Poincaré coordinates, we derive the holographic stress tensor in the back-reacted geometry. The main step is to fix the parameter $\alpha$ for the FG coordinate $z\to z(1-\alpha z^2)$ which reads
\be
\alpha=\frac{\pi^2}{\beta^2}\left(1+\frac{M/R^2}{\left(\cosh\left(\frac{2\pi x_{+}}{\beta}\right)\cosh(\eta^-_{12})-\sinh(\eta^-_{12})\right)\left(\cosh\left(\frac{2\pi x_{-}}{\beta}\right)\cosh(\eta^+_{12})-\sinh(\eta^+_{12})\right)}\right),
\ee
where $\eta^\pm_{12}=\eta_1\pm\eta_2$ and $x_\pm =x\pm t$. Then, the non-zero components of the stress tensor are
\bea
T_{\pm\pm}(x_\pm)=\frac{\pi R}{8G_N \beta^2}+\frac{\pi M}{8G_N \beta^2 R}\frac{1}{\left(\cosh\left(\frac{2\pi x_{\pm}}{\beta}\right)\cosh(\eta^\mp_{12})-\sinh(\eta^{\mp}_{12})\right)^2}\,,\label{eqn:HolStresBTZ}
\eea
reproducing the results in \cite{David:2026owc}. 

%%%%%%%%%%%%%%%%%%%%%%%%%%%%%%%%%%%%%%%%%%%%%%%%
\section{CFT description as local operator quenches}\label{sec:CFTDescription}
%%%%%%%%%%%%%%%%%%%%%%%%%%%%%%%%%%%%%%%%%%%%%%%%
\,\,\,\,\,\,
In this section we consider particular examples of quantum states in 2D CFTs excited by local operators as well as their real time evolution. After defining them, we compute expectation values of the energy-momentum tensors which are universally determined by the operator product expansion (OPE). We also show that, for holographic 2D CFTs, after establishing the appropriate dictionary between the boost and the CFT state parameters, we reproduce the results from the previous section.
%%%%%%%%%%%%%%%%%%%%%%%%%%%%%%%%%%%%%%%%%%%
\subsection{CFT on the plane}
%%%%%%%%%%%%%%%%%%%%%%%%%%%%%%%%%%%%%%%%%%%
\,\,\,\,\,\,
We start with a 2D CFT in Euclidean signature defined on the complex plane with coordinates $(z,\bar{z})=(x+i\tau,x-i\tau)$. Then, consider the following quantum state
\bea
\ket{\psi(t)}&\equiv&e^{-itH}e^{-\frac{\epsilon_++\epsilon_-}{2}H}e^{i\frac{\epsilon_+-\epsilon_-}{2}P}\,O_\Delta(0,0)\ket{0}\,,
\eea
where $\ket{0}$ is the CFT vacuum state and $O_\Delta(0,0)$ is a CFT primary operator with conformal dimension $\Delta=h+\bar{h}=2h$ inserted in spatial position $z=\bar{z}=0$. Operators $H$ and $P$ generate translations in the Euclidean time $\tau$ and space $x$ respectively, and leave the vacuum state invariant. In this way, the initial state is locally excited by the primary operator and regulated/smeared in both the Euclidean time and space by an appropriate amount which depends on the parameters $\epsilon_\pm$.  The state is then time evolved in real, Lorentzian time $t$. In order to perform computations using the Euclidean CFT formalism, we treat $(\epsilon_++\epsilon_-)/2+it$ as a purely real number (Euclidean time) until the end of our computation. This is the standard treatment of quantum quenches in 2D CFTs, due to Calabrese and Cardy \cite{Calabrese:2016xau}. For $\epsilon_+=\epsilon_-=\epsilon$, this state reduces to the standard local operator quench \cite{Nozaki:2014hna,He:2014mwa,Caputa:2014vaa}. Below, we will show that the expectation values of chiral and anti-chiral stress tensors in 2D CFTs can be matched with the holographic stress tensors derived in the previous section.

First, we can write the density matrix corresponding to this state as
\be
\rho(t)=\mathcal{N}\ket{\psi(t)}\bra{\psi(t)}\equiv\mathcal{N}\,O(z_4,\bar{z}_4)\ket{0}\bra{0}O^\dagger(z_1,\bar{z}_1)\,,\label{rhotPlane}
\ee
where $\mathcal{N}$ is the normalization ensuring $\Tr(\rho(t))=1$, and the operator insertion points expressed in complex coordinates are
\bea
&&z_4=-i(\epsilon_-+it)\,,\qquad z_1=i(\epsilon_--it)\,,\nn\\
&&\bar{z}_4=i(\epsilon_++it)\,,\qquad \bar{z}_1=-i(\epsilon_+-it)\,.\label{zsPoincare}
\eea
Observe that, in this notation, the initial density matrix is written as
\be \label{densitymatplane}
\rho(0)= \mathcal{N} O(-i\epsilon_-,i\epsilon_+)\ket{0}\bra{0}O^\dagger(i\epsilon_-,-i\epsilon_+)\,,
\ee
and its finite temperature analogue was discussed in the context of thermal local operator quenches in \cite{David:2026owc}.

Then we evaluate the correlators of the stress tensors as the trace with this density matrix using the universal OPE in 2D CFT. Indeed, introducing $x_\pm=x\pm t$, we simply have 
\bea
T(x_-)&\equiv&\frac{\langle O^\dagger(z_1,\bar{z}_1)T(x)O(z_4,\bar{z}_4)\rangle}{\langle O^\dagger(z_1,\bar{z}_1)O(z_4,\bar{z}_4)\rangle}=\frac{h(z_1-z_4)^2}{(x-z_1)^2(x-z_4)^2}\,,\nn\\
\overline{T}(x_+)&\equiv&\frac{\langle O^\dagger(z_1,\bar{z}_1)\overline{T}(x)O(z_4,\bar{z}_4)\rangle}{\langle O^\dagger(z_1,\bar{z}_1)O(z_4,\bar{z}_4)\rangle}=\frac{\bar{h}(\bar{z}_1-\bar{z}_4)^2}{(x-\bar{z}_1)^2(x-\bar{z}_4)^2}\,.\label{STExpPoinc}
\eea
Moreover, after inserting points \eqref{zsPoincare}, the holographic relation between the CFT operator dimension and the mass of the particle in AdS 
\be
\Delta=2h=mR\,,\label{eqn:mRh}
\ee
and, most importantly:
\be
\epsilon_\pm=R\, e^{\eta_1\pm\eta_2}\,,\label{eq:etaepsilon}
\ee
we reproduce the holographic stress tensors \eqref{HSTPoincare}, and consequently the conserved energy \eqref{TEnP} and momentum \eqref{TPP}\footnote{In both cases taking into account the factor of $-1/(2\pi)$ in the Euclidean CFT definitions of $T$ and $\bar{T}$.}\footnote{It is interesting to note that, from the CFT results: $E=\frac{he^{\eta_2}}{\epsilon}+\frac{he^{-\eta_2}}{\epsilon}$ and $P=\frac{he^{\eta_2}}{\epsilon}-\frac{he^{-\eta_2}}{\epsilon}$, we can see that the unequal smearing of the CFT operators effectively results in unequal scaling dimensions $h\neq\bar{h}$.}.

Let us also make a couple of remarks at this point. First, the CFT results for the expectation values of the stress tensor correlators suggest that we can uniformize the answers in terms of left and right diffeomorphisms. Namely, we can write \eqref{STExpPoinc} as\footnote{Parameter $\alpha$ here should not be confused with $\alpha$ used in the FG coordinates.}
\be
T(x_\pm)=\frac{c}{12}\{f_\pm(x_\pm),x_\pm\}\,,\qquad f_\pm(x_\pm)=\left(\frac{x_\pm-i\epsilon_\pm}{x_\pm+i\epsilon_\pm}\right)^{\alpha}\,,\qquad \alpha=\sqrt{1-\frac{24h}{c}}\,,\label{sqn:TSchwarzain}
\ee
where the Schwarzian derivative is defined as
\be
\{f(z),z\}=f'''/f'-\frac{3}{2}(f''/f')^2\,.\label{ShwarzianDerivative}
\ee
This suggests that, locally, we can treat the back-reacted geometries as Bañados metrics \cite{Banados:1998gg} obtained by maps from Poincaré coordinates by diffeomorphisms $f_\pm(x_\pm)$. Second, for $h=c/32$, the CFT results for the expectation values of stress tensors in local operator quenches reproduce those from a more universal setup of the local quench in CFT \cite{Calabrese:2007mtj,Calabrese:2016xau}. Our setting with non-trivial smeared initial state is a natural generalization of this scenario as well. We will return to this point at the end of the paper.
%%%%%%%%%%%%%%%%%%%%%%%%%%%%%%%%%%%%%%%%%%%
\subsection{CFT on the cylinder}
%%%%%%%%%%%%%%%%%%%%%%%%%%%%%%%%%%%%%%%%%%%
\,\,\,\,\,\,
Next, we generalize the above states to 2D CFTs on the cylinder. More precisely, consider a 2D CFT on the cylinder of circumference $L$, with coordinates $w=\tau+i\sigma$, $\bar{w}=\tau-i\sigma$, with Euclidean time $\tau\in(-\infty,+\infty)$ and periodic spatial coordinate $\sigma\sim\sigma+L$. The cylinder is mapped to the complex plane with coordinates $(z,\bar{z})$ by the exponential maps
\be
z(w)=e^{\frac{2\pi}{L}w}\,,\qquad \bar{z}(\bar{w})=e^{\frac{2\pi}{L}\bar{w}}\,.\label{eq:ExpMapsCylL}
\ee
Then, by analogy with the previous section, we define a density matrix of a pure state created by the real time evolution of the vacuum state on the cylinder excited by a local primary operator
\be
\rho(t)=\mathcal{N}e^{-iHt}O(-\epsilon_-,-\epsilon_+)\ket{0}\bra{0}O^\dagger(\epsilon_-,\epsilon_+)e^{iHt}\equiv \mathcal{N} O(w_4,\bar{w}_4)\ket{0}\bra{0}O^\dagger(w_1,\bar{w}_1)\,,
\ee
with normalization $\mathcal{N}$ ensuring $\Tr(\rho(t))=1$, and the operator's dimension $\Delta=2h$.
In the second equality, using the Euclidean formalism,  we defined
\be
w_4=-\epsilon_-+it,\qquad \bar{w}_4=-\epsilon_++it\,,\qquad w_1=\epsilon_-+it\,,\qquad \bar{w}_1=\epsilon_++it\,.
\ee
Recall that, on the cylinder, the Hamiltonian and momentum operators are given in terms of the global $SL(2,\mathbb{R})$ generators as
\bea
H=\frac{2\pi}{L}\left(L_0+\bar{L}_0-\frac{c}{12}\right)\,,\qquad P=\frac{2\pi}{L}\left(L_0-\bar{L}_0\right)\,,
\eea
so our initial state can be written as
\be
O(-\epsilon_-,-\epsilon_+)\ket{0}=e^{\frac{\epsilon_-+\epsilon_+}{2}H}e^{-\frac{\epsilon_+-\epsilon_-}{2}P}O(0,0)\ket{0}=e^{\frac{2\pi \epsilon_-}{L}L_0 }e^{\frac{2\pi \epsilon_+}{L}\bar{L}_0}O(0,0)\ket{0}\,.
\ee
This way, using different regulators with the Hamiltonian and momentum operators can be interpreted as using different smearing for the left ($L_0$) and right $(\bar{L}_0)$ Hamiltonians on the cylinder\footnote{In the last equation we neglected the constant term proportional to the identity, since it does not affect the dynamics.}.

Next, we can compute the spatial one-point functions of stress tensors 
\bea
T_\mp(x_\mp)\equiv \Tr(\rho(t)T_\mp(\pm i\sigma))&=&\frac{\pi^2 h}{L^2}\frac{\sinh^2\left(\frac{2\pi\epsilon_\mp}{L}\right)}{\sin^2\left(\frac{\pi(x_\mp+i\epsilon_\mp)}{L}\right)\sin^2\left(\frac{\pi(x_\mp-i\epsilon_\mp)}{L}\right)}-\frac{c\pi^2}{6L^2}\,,
\eea
where we denoted $x_\pm=\sigma\pm t$.
Hence, the total energy and momentum expectation values become
\be
E=\frac{2\pi h}{L}\frac{\sinh\left(\frac{2\pi(\epsilon_++\epsilon_-)}{L}\right)}{\sinh\left(\frac{2\pi\epsilon_-}{L}\right)\sinh\left(\frac{2\pi\epsilon_+}{L}\right)}-\frac{c\pi}{6L}\,,\qquad J=\frac{2\pi h}{L}\frac{\sinh\left(\frac{2\pi(\epsilon_+-\epsilon_-)}{L}\right)}{\sinh\left(\frac{2\pi\epsilon_-}{L}\right)\sinh\left(\frac{2\pi\epsilon_+}{L}\right)}\,.
\ee

Finally, comparing with the holographic results \eqref{EJGlobalp} and \eqref{eqn:STGlobalHol}, after using the Brown-Henneaux relation \cite{Brown:1986nw}
\be
c=\frac{3R}{2G_N}\,,\label{eq:Brown-Hen}
\ee
the holographic dictionary between parameters \eqref{eqn:mRh}, and overall $2\pi$ convention for the CFT stress tensor, we find perfect agreement, provided that the relations between the boosts in gravity and the parameters of the CFT state are
\be
\tanh\left(\frac{\pi\epsilon_-}{L}\right)=e^{-(\eta_1+\eta_2)}\,,\qquad\tanh\left(\frac{\pi\epsilon_+}{L}\right)=e^{-(\eta_1-\eta_2)}\,,\label{Globalboosteps}
\ee
or equivalently
\be
e^{-\frac{2\pi\epsilon_-}{L}}=\tanh\left(\frac{\eta_1+\eta_2}{2}\right)\,,\qquad e^{-\frac{2\pi\epsilon_+}{L}}=\tanh\left(\frac{\eta_1-\eta_2}{2}\right)\,.
\ee
This fixes the holographic dictionary between the back-reacted geometry from the massive particle with angular momentum in $AdS_3$ and the generalized local operator quench in 2D CFT on the cylinder.

Let us finally point out that we can also uniformize the stress tensor expectation values above as
\be
T_\pm(x_\pm)=-\frac{c}{12}\{f(x_\pm),x_\pm\}\,,\qquad f_\pm(x_\pm)=\left(\frac{e^{\mp\frac{2\pi ix_\pm}{L}}-e^{-\frac{2\pi \epsilon_\pm}{L}}}{e^{\mp\frac{2\pi ix_\pm}{L}}-e^{\frac{2\pi \epsilon_\pm}{L}}}\right)^{\alpha}\,,
\ee
where $\alpha=\sqrt{1-24h/c}$ as before and, for large $L$, these diffeomorphisms reduce to \eqref{sqn:TSchwarzain}.
%%%%%%%%%%%%%%%%%%%%%%%%%%%%%%%%%%%%%%%%%%%
\subsection{CFT at finite temperature}
%%%%%%%%%%%%%%%%%%%%%%%%%%%%%%%%%%%%%%%%%%%
\,\,\,\,\,\,
Finally, we briefly revisit a CFT on the thermal cylinder with periodic Euclidean time coordinate $\tau\sim\tau+\beta$ and infinite spatial coordinate $x$. The temperature is related to the periodicity of $\tau$ in the usual way, $T=1/\beta$. We will use the complex coordinates on this cylinder $(w,\bar{w})=(x+i\tau,x-i\tau)$ and the maps to the complex plane
\be
z(w)=e^{\frac{2\pi}{\beta}w}\,,\qquad z(\bar{w})=e^{\frac{2\pi}{\beta}\bar{w}}\,.\label{ExpMapsThermalCyl}
\ee
This time, we define a thermal density matrix $\rho_\beta=e^{-\beta H}$ excited by local primary operators with different chiral and anti-chiral regulators \cite{David:2026owc}
\be
\rho(t)=\mathcal{N}e^{-iHt}O(-i\epsilon_-,i\epsilon_+)\rho_\beta O^\dagger(i\epsilon_-,-i\epsilon_+)e^{iHt}\,.
\ee
For $\epsilon_+=\epsilon_-=\epsilon$, these states were studied as local operator quenches at finite temperature in \cite{Caputa:2014eta}.

Using the OPE and exponential maps above, we find that the one-point functions of the stress tensors in this state are expressed in light cone coordinates, $x_\pm=x\pm t$, as
\be
T_{\pm}(x_\pm)\equiv\Tr(\rho(t)T_{\pm\pm}(x))=\frac{\pi^2 h}{\beta^2}\frac{\sin^2\left(\frac{2\pi\epsilon_\pm}{\beta}\right)}{\sinh^2\left(\frac{\pi(x_\pm+i\epsilon_\pm)}{\beta}\right)\sinh^2\left(\frac{\pi(x_\pm-i\epsilon_\pm)}{\beta}\right)}+\frac{c\pi^2}{6\beta^2}\,.
\ee

Comparing with the holographic result \cite{David:2026owc}, reproduced in \eqref{eqn:HolStresBTZ}, we see that the expressions are identical, given the relation between boosts and parameters of the CFT state
\be
e^{-(\eta_1+\eta_2)}=\tan\left(\frac{\pi\epsilon_-}{\beta}\right)\,,\qquad e^{-(\eta_1-\eta_2)}=\tan\left(\frac{\pi\epsilon_+}{\beta}\right)\,.\label{BTZboostepsilon}
\ee

As for the plane and the cylinder, we can write the stress tensors at finite temperature in terms of Schwarzian derivatives of the following functions
\be
f_\pm(x_\pm)=\left(\frac{e^{\frac{2\pi x_\pm}{\beta}}-e^{-\frac{2\pi i \epsilon_\pm}{\beta}}}{e^{\frac{2\pi x_\pm}{\beta}}-e^{\frac{2\pi i\epsilon_\pm}{\beta}}}\right)^{\alpha}\,,\qquad \alpha=\sqrt{1-\frac{24h}{c}}\,.
\ee

This concludes our proposal and matching between the CFT states and the back-reacted metrics in $AdS_3$. In the following section, we will substantiate this dictionary with computations of entanglement entropies in holography and 2D CFTs in the large-$c$ limit.

%%%%%%%%%%%%%%%%%%%%%%%%%%%%%%%%%%%%%%%%%%%%%%%%
\section{Entanglement entropy in $AdS_3/CFT_2$}\label{sec:EntanglementEntropies}
\begin{figure}[h!]
    \centering
    \includegraphics[width=0.9\linewidth]{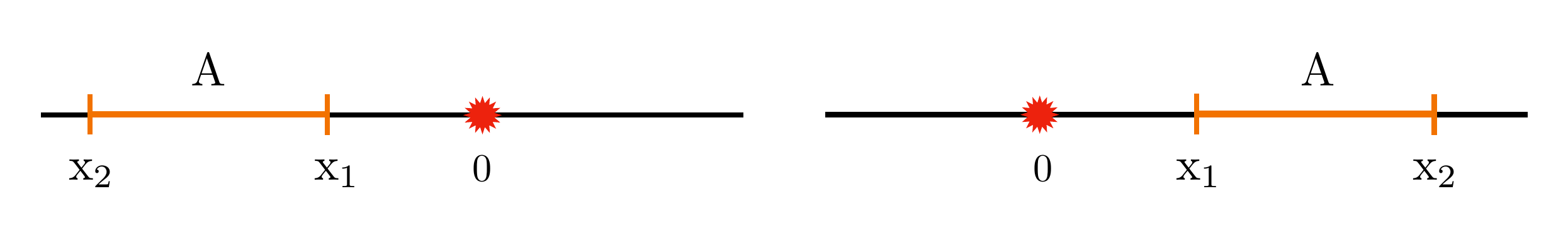}
    \caption{Our setup for computing entanglement entropy for a single interval $A=[x_1,x_2]$, placed to the left or to the right of the excitation at $x=0$.}
    \label{fig:SetupEE}
\end{figure}

%%%%%%%%%%%%%%%%%%%%%%%%%%%%%%%%%%%%%%%%%%%%%%%%
\,\,\,\,\,\,
In the previous sections, we constructed the trajectories of point particles carrying longitudinal velocity or angular momentum in several asymptotically $AdS_3$ spacetimes and analytically derived their back-reacted geometries using coordinate maps involving two independent boost parameters. We then demonstrated that, after an appropriate identification of the bulk and boundary parameters, the holographic stress tensors obtained from the FG expansion precisely reproduce the 2D CFT stress tensors evaluated in states excited by local operators with independent chiral and anti-chiral regulators. In this section, we provide a further non-trivial test of this correspondence by computing the holographic entanglement entropy for a single interval, using the HRT prescription \cite{Hubeny:2007xt} in the back-reacted geometries, and comparing the results with the large-$c$ predictions of the dual CFT. We will consider two cases, when the entangling interval $A$ is chosen to the right and to the left of the local excitation at spatial axis $x=0$ (or $\theta=0$ in global coordinates) on an arbitrary time slice (Fig.\,\ref{fig:SetupEE}).

Our analysis closely follows the framework of holographic local (operator) quenches pioneered in \cite{Nozaki:2013wia}. This line of research has since been developed extensively. In particular, for 2D CFTs, the field-theoretic results of \cite{Nozaki:2014hna,He:2014mwa} were reproduced through exact analytic computations of geodesic lengths in asymptotically $AdS_3$ and BTZ black brane geometries \cite{Caputa:2014eta,Caputa:2015waa,Asplund:2014coa,Asplund:2013zba,David:2017eno}. Since these developments are by now well established, we will keep our discussion brief and refer the reader to the pedagogical review \cite{Calabrese:2016xau} and the textbook \cite{Rangamani:2016dms} for further details.

%%%%%%%%%%%%%%%%%%%%%%%%%%%%%%%%%%%%%%%%%%%%%%%%
\subsection{Holographic entanglement entropy}
%%%%%%%%%%%%%%%%%%%%%%%%%%%%%%%%%%%%%%%%%%%%%%%%
\,\,\,\,\,\,
We begin by computing the holographic entanglement entropy in our time-dependent, back-reacted geometries using the HRT prescription \cite{Hubeny:2007xt}. To derive our main results, we make use of the general expression for the geodesic distance between two points, $(\tau_1,\theta_1,r_1)$ and $(\tau_2,\theta_2,r_2)$, in the back-reacted metric \eqref{BRMetric} in global coordinates. Substituting this expression into the HRT formula, we obtain the following general result for the entanglement entropy:
\begin{align}\label{EEntropy}
    S_A=\frac{c}{6}\log\left[r^\infty_{1}r^\infty_2\,\frac{2\cos\left(\sqrt{1-\frac{M}{R^2}}|\Delta\tau^\infty|\right)-2\cos\left(\sqrt{1-\frac{M}{R^2}}|\Delta\theta^\infty|\right)}{R^2-M}\right]\,,
\end{align}
where the differences between the points are denoted by $|\Delta\theta^\infty|=|\theta_1-\theta_2|$ and $|\Delta\tau^\infty|=|\tau_1-\tau_2|$. To apply this expression to our setup, we use the two-boost maps \eqref{MapGen} and determine the end-points of the entangling interval in global coordinates in \eqref{EEntropy} in terms of the original coordinates describing the spacetime interval at the asymptotic boundary of the back-reacted Poincaré, Global, and BTZ black brane geometries. We analyze each of these cases separately below.
%%%%%%%%%%%%%%%%%%%%%%%%%%%%%%%%%%%%%%%%%%%%%%%%
\subsubsection{Poincaré}
%%%%%%%%%%%%%%%%%%%%%%%%%%%%%%%%%%%%%%%%%%%%%%%%
\,\,\,\,\,\,
The map transforming the massive particle with longitudinal velocity in Poincaré AdS to the static particle at the origin of global AdS was given in \eqref{MapPoincare}. With its help, we can express the boundary points $(\tau_1,r_1,\theta_1)$ and $(\tau_2,r_2,\theta_2)$ in terms of their counterparts in Poincaré coordinates $(t,x,z)$ as
\begin{align}
r_{i}&=\frac{R}{2z\sqrt{\epsilon_-\epsilon_+}} \bigg[x_i^4+x_i^2 \left(\epsilon_+^2-2 t ^2+2 z^2+\epsilon_-
   ^2\right)+2 t  x_i \left(\epsilon_- ^2-\epsilon_+ ^2\right)+z^4\nonumber \\
   &\hspace{1.3cm} +\left(\epsilon_+ ^2+t ^2\right) \left(t ^2+\epsilon_- ^2\right)-2 z^2
   \left(\epsilon_+  \epsilon_- +t ^2\right)\bigg]^{\frac{1}{2}}\,,\nonumber\\
    \tau_{i}&= \tan ^{-1}\left[\frac{t  (\epsilon_+ +\epsilon_- )+x_i (\epsilon_-
   -\epsilon_+ )}{\epsilon_+  \epsilon_- -t ^2+x_i^2+z^2}\right]\,,\\
   \theta_{i}&=\tan ^{-1}\left[\frac{t  (\epsilon_- -\epsilon_+ )+x_i (\epsilon_+
   +\epsilon_- )}{x_i^2-\epsilon_+  \epsilon_- -t ^2+z^2}\right]\,,\nonumber
\end{align}
where $i=1,2$, label $x_1$ and $x_2$ denoting the endpoints of the entangling interval at time t. With a slight abuse of notation, we assumed that the interval is located at some small, radial cut-off position $z$.
In writing the above transformation, we have also used the relation \eqref{eq:etaepsilon}.

Now, we would like to extract the answer in the limit of $\epsilon\to0$ and $v$-fixed\footnote{Recall that $\epsilon$ was the initial location and $v$ the longitudinal velocity of the infalling particle.}. From the relations between the boost parameters and initial conditions \eqref{eq:constr1}, we have $\epsilon_{\pm}=\epsilon e^{\pm\eta_2}$, so this limit requires taking $(\epsilon_-,\epsilon_+)\rightarrow 0$ with their ratio fixed.

\subsection*{Times: $t < |x_1|$ or $t > |x_2|$.}
\,\,\,\,\,\,
Extracting the limit in \eqref{EEntropy}, and retaining terms up to order $\epsilon_-^2$ and $\epsilon_+^2$, we find that for $t < x_1$ and $t > x_2$, the time dependence of the $\epsilon_-^2$ and $\epsilon_+^2$ contributions in the holographic result is
\begin{align}
    \Delta S_A= \frac{mR}{3}\lrt{\epsilon_-^2\frac{(x_1-x_2)^2}{(x_1-t)^2(x_2-t)^2} +\epsilon_+^2 \frac{(x_1-x_2)^2}{(x_1+t)^2(x_2+t)^2}}\,,\label{EEPoincELT}
\end{align}
where in $\Delta S_A$, we always subtract the universal result for the entropy of a single interval, e.g. for the vacuum on the line this contribution is \cite{Holzhey:1994we,Calabrese:2004eu,Ryu:2006bv}
\be 
S_A=\frac{R}{2G_N}\log\left(\frac{|x_1-x_2|}{z}\right)=\frac{c}{3}\log\left(\frac{|x_1-x_2|}{z}\right)\,,\label{hrtbefore}
\ee
where in the second step we used \eqref{eq:Brown-Hen}.

It is interesting to point out the universal nature of the 
terms proportional to $\epsilon_-^2$ and $\epsilon_+^2$ in any 2D CFT, as was found earlier in \cite{david2016universal}.
These terms 
can be understood as a result of the expectation value of the 
stress tensor on the replica surface. 
Let us recall the CFT procedure to   evaluate the change in 
entanglement entropy due to the excited state in (\ref{densitymatplane}). 
We evaluate the  change in R\'{e}nyi entropy 
\begin{eqnarray}
\Delta S_A^{(n)} = \frac{1}{ 1 - n } \log  \frac{{ \Tr} ( \rho_{ A, \epsilon})^n }{\Tr ( \rho_{A} )^n }\,.
\end{eqnarray}
Here $\rho_{A, \epsilon}$ is the reduced density matrix  obtained by tracing outside the interval $(x_1, x_2)$, referred to as $A$ on the density matrix defined in (\ref{densitymatplane}). While  $\rho_{A}$ is the reduced density matrix without any operator insertions.  This factor appears in the denominator  to ensure that we subtract the single interval entanglement entropy of the vacuum. Using the path integral description of the reduced density matrix, the R\'{e}nyi entropy can be evaluated by 
considering the $2n$-point function of the operator ${ O }$ on a  $n$-branched plane. 
The planes are branched over the interval $A$. 
The locations of the operators on the  $j$-th plane  are given  by 
\begin{eqnarray}
 y_1^{(j)} = - i \epsilon_-  , \qquad y_4^{(j)}  = i \epsilon_-, \\ \nonumber
\bar y_1^{(j)} =  i \epsilon_+, \qquad y_4^{(j)} = -i \epsilon_+,
\end{eqnarray}
where $j = 1, 2, \cdots n $. 
The  end points of the interval $A$ in the Lorentzian signature are 
\begin{eqnarray} \label{endpoints}
( y_ 2, \bar y_2 )  = ( x_1 - t, x_1 +t ) , \qquad (y_3, \bar y_3) = ( x_2 - t, x_2+t)\,. 
\end{eqnarray}
For small $\epsilon_\pm$, the operators $O$ and $O^\dagger$ are close to each other in each sheet and are separated 
from the location of the interval. This  is true when $t<x_1$ or $t>x_2$. 
In this regime, the $2n$-point function determining the R\'{e}nyi entropy can be approximated by its OPE expansion 
\begin{eqnarray} \label{ope2n}
& &\prod_{j =1}^n  O^\dagger ( x_4^{(j)} , \bar x_4^{(j)} ) O ( x_1^{(j)}, \bar x_1^{(j)} )  \\ \nonumber
& &\qquad \qquad \qquad \sim \frac{1}{ ( 2 \epsilon_-)^{ 2h n }  ( 2 \epsilon_+)^{ 2 \bar h n }  } 
\left[ 1 -  \sum_{j = 1}^n \left(  \frac{4\epsilon_-^2  h }{c}  T\left( y_1^{(j )}\right )  +  
\frac{4\epsilon_+^2 \bar h  }{c}   \bar T\left( y_1^{(j) }\right)\right)  
\right].
\end{eqnarray}
Here we have assumed that the leading contribution to the OPE is the stress tensor. This assumption 
is true in any CFT in which the lightest operator 
present in the OPE expansion of the operator $O$
is the stress tensor.

We now need to evaluate the expectation value of the stress tensor in the replica geometry. 
For this it is convenient to use the uniformization
 map which maps the branched replica geometry to the plane,
 given by 
 \begin{eqnarray}
 w( y ) = \left(  \frac{y - y_2}{y-y_3} \right)^{\frac{1}{n}} , 
 \qquad 
 \bar w(\bar y )  =  \left(  \frac{\bar y - \bar y_2}{\bar y-\bar y_3} \right)^{\frac{1}{n}}. 
 \end{eqnarray}
The map has branch points at $(y_2, \bar y_2)$ and $(y_3, \bar y_3)$, i.e., the end points of the interval $A$, and these points are mapped to $w=0$ and  $w=\infty$ respectively. One copy of the branched Riemann surface 
is mapped to a wedge in the uniformized plane. 
The expectation value of the stress tensor on the replica geometry  is obtained by its relation to that on the plane
\begin{eqnarray} \label{transstress}
T(y) = w'(y)^2 T( w) + \frac{c}{12} \{ w, y\}\,,
\end{eqnarray}
where the Schwarzian derivative was introduced in \eqref{ShwarzianDerivative}.

Substituting the uniformization map in (\ref{transstress}), we obtain 
\begin{eqnarray}
T(y) &=& \frac{c}{12}  \{ w, y\} = c \frac{ n^2 - 1}{ 24 n^2} \frac{ (y_2 - y_3)^2 }{ ( y - y_2)^2 ( y - y_3)^2}, 
\end{eqnarray}
where we have  used the fact that the expectation value of the stress tensor vanishes on the plane. 
Evaluating the change in the entanglement entropy using the OPE expansion, we find the following leading correction 
\begin{eqnarray}
\Delta S_A^{(n)} &=&  \frac{c ( n +1) }{ 6  n}  \left ( h \epsilon_-^2 \frac{ ( y_2- y_3)^2}{ y_2^2 y_3^2 } 
+ \bar h \epsilon_+^2  \frac{ ( \bar y_2- \bar y_3)^2}{ \bar y_2^2  \bar y_3^2 }  \right).
\end{eqnarray}
Note that the  density matrix  in (\ref{densitymatplane})
is normalized using the 2 point function of the operators on the plane.  The normalization $\mathcal{N}$
removes the prefactor  in (\ref{ope2n}). We have also  performed the sum over the replica images and kept the leading contribution on substituting for the positions $y_1 ^{(i)}, y_4^{(i)} $. For the infalling conical defect from  (\ref{eqn:mRh}), we have 
\begin{eqnarray}
h = \bar h =  m R.
\end{eqnarray}
Using this, 
and the end points of the interval  $y_2$, $y_3$ from (\ref{endpoints}), we obtain 
\begin{eqnarray}
\lim_{n\rightarrow 1} \Delta S_A^{(n)} = \frac{mR}{3}\lrt{\epsilon_-^2\frac{(x_1-x_2)^2}{(x_1-t)^2(x_2-t)^2} +\epsilon_+^2 \frac{(x_1-x_2)^2}{(x_1+t)^2(x_2+t)^2}}\,,
\end{eqnarray}
which precisely agrees with the result (\ref{hrtbefore}) obtained using the HRT prescription in the geometry of the in-falling conical defect.

\subsection*{Times: $|x_1| < t < |x_2|$.}
\,\,\,\,\,\,
The behaviour near the endpoints of the interval, as well as in the intermediate-time regime $|x_1| < t < |x_2|$, is more subtle and can be determined by carefully analyzing the holographic formula above. For these times, when the ``energy pulse" lies inside the interval, placed either on the left or on the right of the initial excitation (see Fig.\,\ref{fig:SetupEE}), the coordinate transformation must be modified. Since the $\tau$ and $\theta$ coordinates involve inverse tangent functions, two of them are shifted by a phase of $\pi$, leading to a jump in the entanglement entropy. This gives the increase in the entropy for the interval on the right ($0<x_1<x_2$)
\begin{align}
    \Delta S_A&=\frac{c}{6}\log\lrt{\frac{1}{\epsilon_-}\frac{\sin\lr{\pi\alpha}}{\alpha}\frac{(x_1-t)(x_2-t)}{(x_1-x_2)}} - \epsilon_-\lrt{\frac{c}{6}\frac{\alpha}{\tan{\pi\alpha}} \frac{(x_1-x_2)}{(x_1-t)(x_2-t)}}\,,\label{EntPoInsideInt}
\end{align}
where we already replaced $G_N$ by $c$ with \eqref{eq:Brown-Hen} and denoted 
\be
\alpha=\sqrt{1-\frac{M}{R^2}}=\sqrt{1-8mG_N}\,,\label{eq:alphaGR}
\ee
for comparisons with CFT results.

\begin{figure}[h]
    \begin{subfigure}{0.46\linewidth}
     \centering
 \includegraphics[width=.8\linewidth]{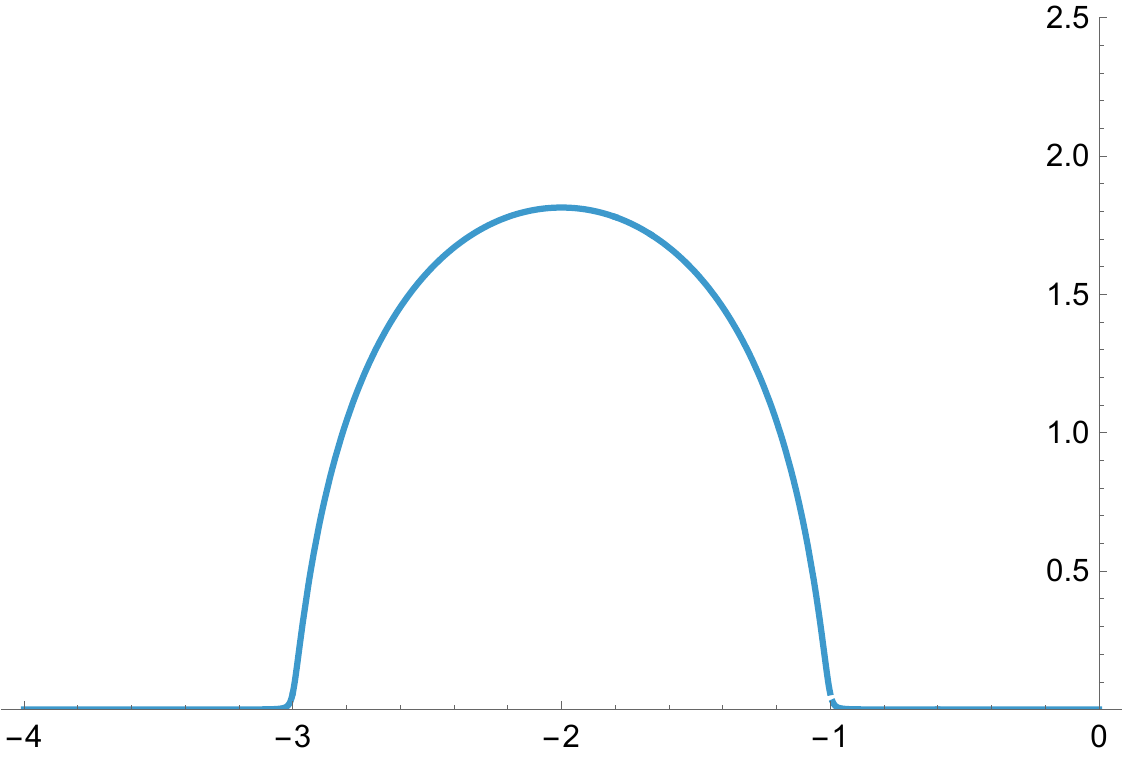}
 \caption{$\Delta S_A$ for the interval on the left.}
 \label{fig 3 a}
 \end{subfigure}\hfill
 \begin{subfigure}{0.46\linewidth}
  \centering
 \includegraphics[width=.8\linewidth]{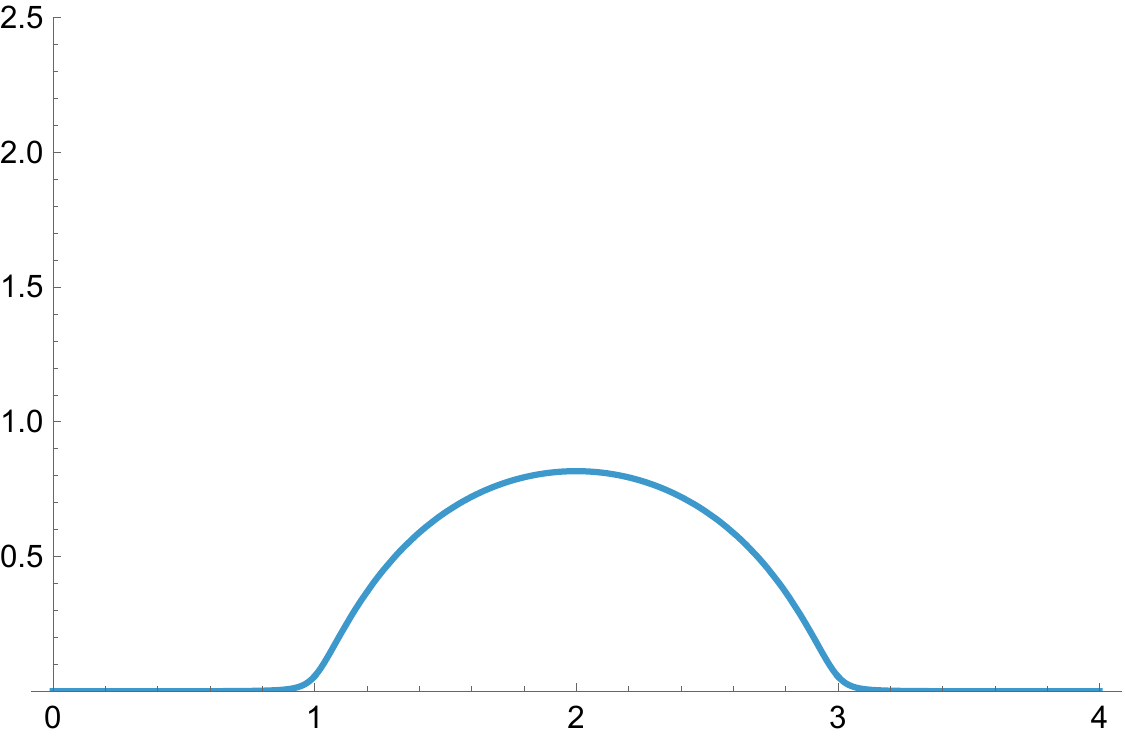}
 \caption{$\Delta S_A$ for the interval on the right.}
\label{fig 3 b}
\end{subfigure}
    \caption{Time evolution of the increase in entanglement entropy $\Delta S_A$. Both plots are generated for $\epsilon_-=0.08,\ \epsilon_+=0.02$, and $\frac{2mG_N}{3}=\frac{\Delta}{c}= \text{fixed}$.}
    \label{fig:current}
\end{figure}

Similarly, for the entangling interval placed to the left of the excitation $(x_2<x_1<0)$, we find 
\begin{align}
    \Delta S_A&=\frac{c}{{6}}\log\lrt{\frac{1}{\epsilon_+}\frac{\sin\lr{\pi\alpha}}{\alpha}\frac{(x_1+t)(x_2+t)}{(x_2-x_1)}}+ \epsilon_+\lrt{\frac{c}{6}\frac{\alpha}{\tan{\pi\alpha}} \frac{(x_1-x_2)}{(x_1+t)(x_2+t)}}\,.
\end{align}
It is clear that the dynamics of entanglement in this local operator quench setup shows asymmetry between the left and right moving parts of the excitations (using the intuitive quasi-particle picture). Indeed, as can be seen in Fig.\,\ref{fig:current}, the right movers and their height are determined by $\epsilon_-=\epsilon e^{-\eta_2}$ and the left movers by $\epsilon_+=\epsilon e^{\eta_2}$. When $v=0$ ($\eta_2=0$) the pulses get equal heights and reproduce the original results \cite{Nozaki:2014hna}.

In the next section we will compare these results with explicit computations in large-$c$ 2D CFTs.
%%%%%%%%%%%%%%%%%%%%%%%%%%%%%%%%%%%%%%%%%%%%%%%%
\subsubsection{Global AdS}
%%%%%%%%%%%%%%%%%%%%%%%%%%%%%%%%%%%%%%%%%%%%%%%%
\,\,\,\,\,\,
We now move to the computation of entanglement entropy in back-reacted global AdS spacetime. The asymptotic boundary is now the cylinder and, analogously to Fig.\,\ref{fig:SetupEE}, we choose the entangling interval as $A=[\phi_1,\phi_2]$ at time $t$, to the left or to the right of the local excitation at $\phi=0$.

The map between the boundary points in this case is given by
\begin{align}
    r_i&=\frac{R}{4\pi^2Lz\sqrt{\epsilon_-\epsilon_+}}\ltb{\left\{\left(L^2+\pi ^2 z^2\right) \sqrt{L^2-4 \pi ^2 \epsilon_+  \epsilon_- }
   \cos \left(\frac{2 \pi t }{L}\right)+L \left(\pi ^2 z^2-L^2\right)
   \cos \left(\frac{2 \pi  \phi_i}{L}\right)\right\}^2\nonumber\\
   &+\pi ^2 \left\{(\epsilon_- -\epsilon_+ ) \left(L^2+\pi ^2 z^2\right) \sin
   \left(\frac{2 \pi t }{L}\right)+(\epsilon_+ +\epsilon_- ) (L^2-\pi^2  z^2)
    \sin \left(\frac{2 \pi  \phi_i}{L}\right)\right\}^2}^{\frac{1}{2}},\nonumber\\ 
\tau_i&=\tan^{-1}\lrt{\frac{\pi  (\epsilon_+ -\epsilon_- ) (\pi^2 z^2-L^2) \sin \left(\frac{2 \pi 
   \phi _i}{L}\right)+\pi  \left(\epsilon _-+\epsilon _+\right)
   \left(L^2+\pi ^2 z^2\right) \sin \left(\frac{2 \pi  t}{L}\right)}{\left(\pi ^2 z^2-L^2\right) \sqrt{L^2-4 \pi ^2 \epsilon _- \epsilon _+}
   \cos \left(\frac{2 \pi  \phi _i}{L}\right)+L \left(L^2+\pi ^2
   z^2\right) \cos \left(\frac{2 \pi  t}{L}\right)}},\\
   \theta_i&=\tan^{-1}\lrt{\frac{\pi  \left(\epsilon _-+\epsilon _+\right) (\pi^2 z^2-L^2) \sin
   \left(\frac{2 \pi  \phi _i}{L}\right)+\pi  \left(\epsilon _+-\epsilon
   _-\right) \left(L^2+\pi ^2 z^2\right) \sin \left(\frac{2 \pi 
   t}{L}\right)}{\left(L^2+\pi ^2 z^2\right) \sqrt{L^2-4 \pi ^2 \epsilon_-  \epsilon_+ } \cos
   \left(\frac{2 \pi  \tau }{L}\right)+L \left(\pi ^2 z^2-L^2\right) \cos
   \left(\frac{2 \pi \phi_i}{L}\right)}}\nonumber,
\end{align}
where we used the cut-off $z$ related to $\rho$ in \eqref{MAPGLOBALC} by $\rho= \log{\frac{L}{\pi z}}$, 
and the boost parameters have been replaced using the relations \eqref{Globalboosteps}.

The entanglement entropy can be computed by substituting the above transformation in \eqref{EEntropy}. Its evolution is very similar to that of the Poincaré cases. In the early time regime $t<|\phi_{1,2}|$ as well as the late time regime $t>|\phi_{1,2}|$, and to the leading orders in $\epsilon_\pm$, the increase in entanglement entropy evolves as 
\begin{align} \label{epsilonsqcy}
    \Delta S_A%&=\frac{c}{6}\log\lrt{\frac{L^2}{\pi^2z^2}\sin^2{\frac{\pi}{L}(\phi_1-\phi_2)^2}}\nonumber\\
    =  \frac{mR}{3}\lrt{\epsilon_-^2\frac{\pi^2}{L^2}\frac{\sin^2\frac{\pi}{L}(\phi_1-\phi_2)}{\sin^2\frac{\pi}{L}(\phi_1-t) \sin^2\frac{\pi}{L}(\phi_2-t)} + \epsilon_+^2\frac{\pi^2}{L^2}\frac{\sin^2\frac{\pi}{L}(\phi_1-\phi_2)}{\sin^2\frac{\pi}{L}(\phi_1+t) \sin^2\frac{\pi}{L}(\phi_2+t)}},
\end{align}
where we subtracted the single interval contribution in the finite size-L setting
\be
S_A=\frac{c}{3}\log\left[\frac{L}{\pi z}\sin\frac{\pi|\phi_1-\phi_2|}{L}\right]\,.
\ee
Again, the correction proportional to $\epsilon_-^2, \epsilon_+^2$,  arises due to the expectation value of the 
stress tensor on the replica geometry. 
But in this case the CFT is on the branched cylinder and therefore the uniformization  map that we need to employ 
is given by 
\begin{eqnarray}
    w(y) = e^{\frac{i \pi(y_2 -y_3)}{n L}}
    \left[\frac{\sin\frac{\pi}{L}(y-y_2)}{\sin\frac{\pi}{L}( y - y_3) } \right]^{\frac{1}{n}}, \\ \nonumber
    \bar w(\bar y) = e^{\frac{i \pi(\bar y_2 -\bar y_3)}{n L }}
    \left[\frac{\sin\frac{\pi}{L }(\bar y- \bar y_2)}{\sin\frac{\pi}{L }( \bar y - \bar y_3) } \right]^{\frac{1}{n}}.
\end{eqnarray}
Following the same steps as discussed for the uniformization map from the branched plane we obtain the result in (\ref{epsilonsqcy}) for the $\epsilon^2$ correction when the pulse is outside the entangling interval.

On the other hand, in the intermediate time regime $\phi_1<t<\phi_2$, we again derive an increase in the entropy with the leading-order correction for the interval on the right
\begin{align}
  \Delta S_A&=\frac{c}{6}\ \log{\lr{\frac{1}{\epsilon_-}\frac{L}{\pi}\frac{\sin{\lr{\pi\alpha}}}{\alpha}\frac{\sin\frac{\pi}{L}(\phi_1-t)\sin\frac{\pi}{L}(\phi_2-t)}{\sin\frac{\pi}{L}(\phi_1-\phi_2)}}}\nonumber \\
    &-\epsilon_-\lrt{\frac{c}{6}\frac{\pi}{L}\frac{\alpha}{\tan{\lr{\pi\alpha}}}\frac{\sin\frac{\pi}{L}(\phi_1-\phi_2)}{\sin\frac{\pi}{L}(\phi_1-t)\sin\frac{\pi}{L}(\phi_2-t)}} \,,
\end{align}
and for the interval on the left of the excitation
\begin{align}
  \Delta S_A&= \frac{c}{6}\ \log{\lrt{\frac{1}{\epsilon_+}\frac{L}{\pi}\frac{\sin{\lr{\pi\alpha}}}{\alpha}\frac{\sin\frac{\pi}{L}(\phi_1+t)\sin\frac{\pi}{L}(\phi_2+t)}{\sin\frac{\pi}{L}(\phi_1-\phi_2)}}}\nonumber \\
    &-\epsilon_+\lrt{\frac{c}{6}\frac{\pi}{L}\frac{\alpha}{\tan{\lr{\pi\alpha}}}\frac{\sin\frac{\pi}{L}(\phi_1-\phi_2)}{\sin\frac{\pi}{L}(\phi_1+t)\sin\frac{\pi}{L}(\phi_2+t)}}\,,
\end{align}
where in both cases we already employed \eqref{eq:Brown-Hen} to replace $G_N$ in the HRT formulas by $c$ and used \eqref{eq:alphaGR}.

Similarly to the infinite size (recovered by taking $L\to\infty$) we observe two pulses of different magnitudes determined by $\epsilon_\pm$. The answers are of course periodic in time and we will discuss them in more detail when comparing with 2D CFTs at large central charge.
%%%%%%%%%%%%%%%%%%%%%%%%%%%%%%%%%%%%%%%%%%%%%%%%
\subsubsection{BTZ}
%%%%%%%%%%%%%%%%%%%%%%%%%%%%%%%%%%%%%%%%%%%%%%%%
\,\,\,\,\,\,
Finally, we compute the entanglement entropy in the back-reacted BTZ black string with coordinates $(t,x,z)$. 
For the HRT, we want to find the geodesic length between points $(t,x_1,z)$ and $(t,x_2,z)$ and we get it by using \eqref{EEntropy} and our general map with two boosts \label{MapGen1} adapted to the BTZ black string coordinates \eqref{MapPSBTZ}
\begin{align}
r_i &=
\frac{R}{M^2z\sqrt{\epsilon_-\epsilon_+}}
\Bigg[
\frac{M^2(\epsilon_-+\epsilon_+)^2}{4}
\left(
\sinh Mx_i
+\frac{\epsilon_- -\epsilon_+}{\epsilon_-+\epsilon_+}\sinh Mt
\right)^2
\nonumber\\
&+\left(
\cosh Mt
-\sqrt{1-M^2\epsilon_-\epsilon_+}\,
\cosh Mx_i
\right)^2
\Bigg]^{1/2},
\nonumber\\[1ex]
\tan\tau_i
&=
\frac{M(\epsilon_-+\epsilon_+)}{2}
\frac{
\sinh Mt
+\frac{\epsilon_- -\epsilon_+}{\epsilon_-+\epsilon_+}\sinh Mx_i
}{
\cosh Mx_i
-\sqrt{1-M^2\epsilon_-\epsilon_+}\cosh Mt
}\,,
\\
\tan\theta_i
&=
\frac{M(\epsilon_-+\epsilon_+)}{2}
\frac{
\sinh Mx_i
+\frac{\epsilon_- -\epsilon_+}{\epsilon_-+\epsilon_+}\sinh Mt
}{
\cosh Mt
-\sqrt{1-M^2\epsilon_-\epsilon_+}\cosh Mx_i
}\,,
\nonumber
\end{align}
where we used the relation \eqref{BTZboostepsilon} between the boosts and $\epsilon_\pm$.

Following exactly the same steps as in the two examples above, the leading order in $\epsilon_-,\epsilon_+$ expression in the early $t<|x_{1,2}|$ and late $t>|x_{1,2}|$ times is
\begin{align}\label{epsilonsqtemp}
    \Delta S_A&= \frac{\Delta}{3}\lrt{ \epsilon_-^2\frac{\pi^2}{\beta^2} \frac{\sinh^2{\frac{\pi}{\beta}(x_1-x_2)}}{\sinh^2{\frac{\pi}{\beta}(x_1-t)}\sinh^2{\frac{\pi}{\beta}(x_2-t)}} + \epsilon_+^2 \frac{\pi^2}{\beta^2}\frac{\sinh^2{\frac{\pi}{\beta}(x_1-x_2)}}{\sinh^2{\frac{\pi}{\beta}(x_1+t)}\sinh^2{\frac{\pi}{\beta}(x_2+t)}}}\nonumber\\
    &\equiv \frac{\Delta}{3}\lrt{\epsilon_-^2 S^2_{x_1x_2}(t) + \epsilon_+^2 S^2_{x_1x_2}(-t) }\,,
\end{align}
where this time we subtracted the finite temperature answer for the single interval entanglement entropy
\be
S_A=\frac{c}{3}\log\lr{\frac{\beta}{z\pi}\sinh\frac{\pi|x_1-x_2|}{\beta}}\,.
\ee
The terms proportional to $\epsilon_\pm^2$ can be understood 
in terms of the expectation value of the stress tensor which 
arises from the OPE just as in the case of the 
infalling particle in Poincar\'{e} $AdS$. 
However in this case we use the map from the branched thermal cylinder to the  uniformized plane which is given by 
\begin{eqnarray}
    w(y) = e^{\frac{\pi(y_2 -y_3)}{n\beta}}
    \left[\frac{\sinh\frac{\pi}{\beta}(y-y_2)}{\sinh\frac{\pi}{\beta}( y - y_3) } \right]^{\frac{1}{n}}, \\ \nonumber
    \bar w(\bar y) = e^{\frac{\pi(\bar y_2 -\bar y_3)}{n\beta}}
    \left[\frac{\sinh\frac{\pi}{\beta}(\bar y- \bar y_2)}{\sinh\frac{\pi}{\beta}( \bar y - \bar y_3) } \right]^{\frac{1}{n}}.
\end{eqnarray}
Then following the same analysis done for the uniformization
map from the branched plane, we obtain the result in 
(\ref{epsilonsqtemp}).

When the pulse enters the interval, the change in entanglement entropy for the interval on the left and the right (as in Fig.\,\ref{fig:SetupEE})  is controlled by $\epsilon_+$ and $\epsilon_-$, respectively. Technically, this arises by choosing different phases in the coordinate transformation. For the interval to the right of the initial excitation, we derive
\begin{align}\label{DSABTZRIGHT}
    \Delta S_A&= \frac{c}{6}\ \log{\lrt{\frac{1}{\epsilon_-}\frac{\beta}{\pi}\frac{\sin{({\pi\alpha})}}{\alpha}\frac{\sinh{\frac{\pi}{\beta}(x_1-t)}\sinh{\frac{\pi}{\beta}(x_2-t)}}{\sinh{\frac{\pi}{\beta}(x_1-x_2)}}}}\nonumber \\
    &-\epsilon_-\lrt{\frac{c}{6}\frac{\pi}{\beta}\frac{\alpha}{\tan{({\pi\alpha})}}\frac{\sinh{\frac{\pi}{\beta}(x_1-x_2)}}{\sinh{\frac{\pi}{\beta}(x_1-t)}\sinh{\frac{\pi}{\beta}(x_2-t)}}}\,.
\end{align}
Similarly, for the interval to the left of the excitation, the increase in entanglement entropy becomes
\begin{align} 
    \Delta S_A&=\frac{c}{6}\ \log{\lrt{\frac{1}{\epsilon_+}\frac{\beta}{\pi}\frac{\sin{\lr{\pi\alpha}}}{\alpha}\frac{\sinh{\frac{\pi}{\beta}(x_1+t)}\sinh{\frac{\pi}{\beta}(x_2+t)}}{\sinh{\frac{\pi}{\beta}(x_2-x_1)}}}}\nonumber \\
    &-\epsilon_+\lrt{\frac{c}{6}\frac{\pi}{\beta}\frac{\alpha}{\tan{\lr{\pi\alpha}}}\frac{\sinh{\frac{\pi}{\beta}(x_2-x_1)}}{\sinh{\frac{\pi}{\beta}(x_1+t)}\sinh{\frac{\pi}{\beta}(x_2+t)}}}\,,\label{DSABTZLEFT}
\end{align}
and in both expressions we used \eqref{eq:Brown-Hen} and parameter $\alpha$ defined in \eqref{eq:alphaGR}.

In the following subsection, we will reproduce these results in 2D CFTs in the large-$c$ approximation.
%%%%%%%%%%%%%%%%%%%%%%%%%%%%%%%%%%%%%%%%%%%%%%%%
\subsection{Entanglement entropy in large-$c$ CFTs}
%%%%%%%%%%%%%%%%%%%%%%%%%%%%%%%%%%%%%%%%%%%%%%%%
\,\,\,\,\,\,
We now turn to the computation of entanglement entropy in 2D CFTs in the large-$c$ approximation and derive the entanglement entropy for a single interval in the time-evolved local operator quenches introduced in Sec.~\ref{sec:CFTDescription}.

Recall that the entanglement entropy of a state $\ket{\psi}$ in a 2D CFT can be computed using the replica trick \cite{Calabrese:2004eu}. This reduces the calculation to evaluating correlation functions of twist operators inserted at the endpoints of the entangling interval $A$. More specifically, the trace of the $n$-th power of the reduced density matrix, $\rho_A=\Tr_{A^c}(\rho)$, is given by the two-point function of the local twist operators $\sigma_n$ (in the $n$-fold orbifold theory) whose conformal dimension is
\be
h_\sigma=\frac{c}{24}(n-1/n)\,,
\ee
inserted at the end points of the interval $A$ (we denoted its complement as $A^c$ and assumed the Hilbert space $\mathcal{H}=\mathcal{H}_A\otimes\mathcal{H}_{A^c}$). For our local operator quenches \eqref{rhotPlane}, this becomes
\be
\Tr\rho^n_A(t)=\frac{\langle O^\dagger(z_1,\bar{z}_1)\sigma_n(z_2,\bar{z}_2)\tilde{\sigma}_n(z_3,\bar{z}_3)O(z_4,\bar{z}_4)\rangle}{\langle O^\dagger(z_1,\bar{z}_1)O(z_4,\bar{z}_4)\rangle^n_1}\,,\label{C4Plane}
\ee
where the complex coordinates for the interval $A=[x_1,x_2]$ are $z_2=\bar{z}_2=x_1$ and $z_3=\bar{z}_3=x_2$, and points $z_1$ and $z_4$ were given in \eqref{zsPoincare} for the line and similarly in later subsections for CFTs on the cylinder and at finite temperature.

In general, this four-point correlator is not universal and depends on the specific 2D CFT under consideration. Namely, its form depends on both, the central charge and the full operator spectrum. However, to reproduce the holographic results of the previous section, it is sufficient to impose the standard constraints expected for holographic 2D CFTs, namely the large-$c$ limit together with a sparse light spectrum \cite{Hartman:2014oaa}. Under these assumptions, further progress can be made. Indeed, if the conformal dimension of the operator creating the excitation is also taken to be large, with $\Delta/c$ held fixed as $c\to\infty$, the relevant replica four-point function can be evaluated using the Heavy-Heavy-Light-Light (HHLL) approximation \cite{Fitzpatrick:2014vua}. In this limit, we get that the vacuum conformal block dominates and the 4-point correlator can be written as
\begin{align}
&\langle O^H(z_1,\bar{z}_1) O^L(z_2,\bar{z}_2)O^L(z_3,\bar{z}_3)O^H(z_4,\bar{z}_4)\rangle=|z_{14}|^{-4h_H}|z_{23}|^{-4h_L}\nonumber\\
&\times \left(\frac{z^{\frac{1-\alpha_+}{2}}(1-z^{\alpha_+})}{\alpha_+ (1-z)}\right)^{-4h_L}\left(\frac{\bar{z}^{\frac{1-\alpha_-}{2}}\left(1-\bar{z}^{\alpha_-}\right)}{\alpha_- (1-\bar{z})}\right)^{-4h_L}\,,
\end{align}
where we denoted the dimensions of the heavy and light operators by $h_H$ and $h_L$ respectively, and parameters $\alpha_\pm$ are given by
\be
\alpha_+=\sqrt{1-\frac{24h_H}{c}},\qquad \alpha_-=\sqrt{1-\frac{24\bar{h}_H}{c}}\,.
\ee
Finally, we have $z_{ij}\equiv z_i-z_j$ and the conformal cross-ratios are
\be
z=\frac{z_{12}z_{34}}{z_{13}z_{24}}\,,\qquad \bar{z}=\frac{\bar{z}_{12}\bar{z}_{34}}{\bar{z}_{13}\bar{z}_{24}}\,, 
\ee
and, during Lorentzian evolution, can be treated as independent variables.

In the remainder of this section, we focus on the entanglement entropy, obtained by taking the replica limit $n\to 1$. In this limit, the twist operators become light, allowing us to treat them as the ``Light'' operators in the HHLL correlators ($h_L=h_\sigma$) and apply the above results to each of our three examples. In this approximation, the increase in entanglement entropy in the heavy operator quench can be simply expressed as \cite{Asplund:2014coa}
\be
\Delta S_A=\frac{c}{6}\log\left[\frac{z^{\frac{1-\alpha_+}{2}}(1-z^{\alpha_+})}{\alpha_+(1-z)}\frac{\bar{z}^{\frac{1-\alpha_-}{2}}(1-\bar{z}^{\alpha_-})}{\alpha_-(1-\bar{z})}\right]\,.\label{DSAHHL}
\ee
This is our final expression, which we will evaluate below for different cross-ratios that change due to the conformal maps from the plane to the cylinders.

%%%%%%%%%%%%%%%%%%%%%%%%%%%%%%%%%%%%%%%%%%%%%%%%
\subsubsection{CFT on the plane}
%%%%%%%%%%%%%%%%%%%%%%%%%%%%%%%%%%%%%%%%%%%%%%%%
\,\,\,\,\,\,
We start with the heavy local operator excitation of the vacuum state of a 2D CFT on the infinite line. We choose the entangling interval $A=|x_2-x_1|$ to have size $l$ so $|x_2|=|x_1|+l$. For the interval on the right ($x_2>x_1>0$), inserting the points \eqref{zsPoincare} the conformal cross-ratios in the small $\epsilon_\pm$ expansion become
\bea
z&=&\frac{(t-x_1+i\epsilon_-)(x_1+l-t+i\epsilon_-)}{(t-x_1-i\epsilon_-)(x_1+l-t-i\epsilon_-)}\simeq1-\frac{2il\epsilon_-}{(x_1-t)(x_1+l-t)}\,,\nn\\
\bar{z}&=&\frac{(x_1+t+i\epsilon_+)(x_1+l+t-i\epsilon_+)}{(x_1+t-i\epsilon_+)(x_1+l+t+i\epsilon_+)}\simeq 1+\frac{2il\epsilon_+}{(x_1+t)(x_1+l+t)}.
\eea
This implies that for $t<x_1$ and $t>x_1+l$ both $z,\bar{z}\to1$ and, to the leading order in $\epsilon_\pm$, $\Delta S_A=0$ and the sub-leading order perfectly reproduces \eqref{EEPoincELT} with $mR=\Delta=2h_H=2\bar{h}_H$. We have already discussed the universality of these leading order corrections (in any 2D CFT) from the OPE perspective during holographic computations in the previous subsection.

On the other hand, for $x_1<t<x_1+l$ the imaginary part of $z$ changes the sign, so we first take $z\to e^{-2\pi i}z$ and then $z,\bar{z}\to1$ obtaining
\be
\Delta S_A=\frac{c}{6}\log\left[\frac{\sin(\pi \alpha_+)}{\alpha_+}\frac{(x_1+l-t)(t-x_1)}{l\epsilon_-}\right]+O(\epsilon_-).
\ee
In fact, we can also carefully evaluate the first $O(\epsilon_-)$ correction and confirm that it perfectly matches \eqref{EntPoInsideInt}\footnote{Reproducing this sub-leading term is a bit subtle when combined with the monodromy. The correct prescription is to take $z\to e^{-2\pi i} z$, followed by $z\simeq e^{i\delta}$, with $\delta=\frac{2l\epsilon_-}{(t-x_1)(x_1+l-t)}$, and then expand for small $\epsilon_-$. }.

Analogous results can be derived if we place the entangling interval to the left of the excitation, and we get the cross-ratios from the ones above by replacing $(x_1,x_2)\to (-x_1,-x_2)$. This time, for $|x_1|<t<|x_1|+l$ we have $\bar{z}\to e^{-2\pi i}\bar{z}$ and $z,\bar{z}\to1$, which yields
\be
\Delta S_A=\frac{c}{6}\log\left[\frac{\sin(\pi \alpha_-)}{\alpha_-}\frac{(|x_2|-t)(t-|x_1|)}{l\epsilon_+}\right]+O(\epsilon_+).
\ee
Once again, we find perfect agreement with the holographic computation with $h_H=\bar{h}_H$, including the sub-leading corrections. This provides further evidence for our proposed correspondence between point particles with non-zero longitudinal velocity in the bulk and heavy local operators smeared with both the Hamiltonian and momentum operators in the dual holographic CFT. We now extend this analysis to the cases of a finite-size CFT and a CFT at finite temperature.
%%%%%%%%%%%%%%%%%%%%%%%%%%%%%%%%%%%%%%%%%%%%%%%%
\subsubsection{CFT on the cylinder}
%%%%%%%%%%%%%%%%%%%%%%%%%%%%%%%%%%%%%%%%%%%%%%%%
\,\,\,\,\,\,
A very similar computation can be done in CFT on a circle of size $L$. Taking the interval to be $A=[\sigma_1,\sigma_2=\sigma_1+l]$ on the spatial circle $\sigma\sim \sigma+L$, and using the exponential map \eqref{eq:ExpMapsCylL}, the cross ratios become
\be
z=\frac{\sin\left(\frac{\pi(\sigma_1+l-t+i\epsilon_-)}{L}\right)\sin\left(\frac{\pi(\sigma_1-t-i\epsilon_-)}{L}\right)}{\sin\left(\frac{\pi(\sigma_1+l-t-i\epsilon_-)}{L}\right)\sin\left(\frac{\pi(\sigma_1-t+i\epsilon_-)}{L}\right)}\simeq 1-\frac{2\pi i\epsilon_- \sin\left(\frac{\pi l}{L}\right)}{L\sin\left(\frac{\pi (\sigma_1-t)}{L}\right)\sin\left(\frac{\pi (\sigma_1+l-t)}{L}\right)}+O(\epsilon^2_-)\,,
\ee
\be
\bar{z}=\frac{\sin\left(\frac{\pi(\sigma_1+l+t-i\epsilon_+)}{L}\right)\sin\left(\frac{\pi(\sigma_1+t+i\epsilon_+)}{L}\right)}{\sin\left(\frac{\pi(\sigma_1+l+t+i\epsilon_+)}{L}\right)\sin\left(\frac{\pi(\sigma_1+t-i\epsilon_+)}{L}\right)}\simeq 1+\frac{2\pi i\epsilon_+ \sin\left(\frac{\pi l}{L}\right)}{L\sin\left(\frac{\pi (\sigma_1+t)}{L}\right)\sin\left(\frac{\pi (\sigma_1+l+t)}{L}\right)}+O(\epsilon^2_+)\,.
\ee
This implies that the imaginary part of the chiral contribution changes sign in the time interval $\sigma_1+nL<t<\sigma_1+l+nL$, where $n=0,1,\ldots$. Similarly, the imaginary part of $\bar z$ changes sign for $L-(\sigma_1+l)+nL<t<L-\sigma_1+nL$. This behaviour is precisely what is expected from the quasiparticle picture: the entanglement entropy receives a non-trivial contribution whenever exactly one of the two members of the EPR pair (the left- or right-moving quasiparticle, which can be intuitively identified with the profiles of the energy density) lies inside the interval $A$, while the other remains outside. Outside these time windows (and assuming, for simplicity, that they do not overlap, so that both quasiparticles are simultaneously inside the interval) we have $(z,\bar{z})\sim (1,1)$ and $\Delta S_A=0$ to the leading order in small $\epsilon_\pm$, and we find that the sub-leading quadratic corrections with $h_H=\bar{h}_H$ match those in the holographic computation. 

Now, the non-trivial contributions to entanglement entropy are obtained in two scenarios. First, when $\sigma_1+nL<t<\sigma_1+l+nL$ we have $z\sim e^{-2\pi i}$ $(\bar{z}\to 1)$ such that
\be
\Delta S_A\simeq\frac{c}{6}\log\left[\frac{\sin(\pi\alpha_+)}{\alpha_+}\frac{L}{\pi \epsilon_-}\frac{\sin\left(\frac{\pi(\sigma_1+l-t)}{L}\right)\sin\left(\frac{\pi(t-\sigma_1)}{L}\right)}{\sin\left(\frac{\pi l}{L}\right)}\right]\,.\label{Window1}
\ee
Then, for the second time window we discussed above, we can simply set $t=L-(\sigma_1+l)+t'$, $0<t'<l$ and derive $\bar{z}\to e^{2\pi i}$, $(z\to1)$ such that
\be
\Delta S_A=\frac{c}{6}\log\left[\frac{\sin(\pi\alpha_-)}{\alpha_-}\frac{L}{\pi \epsilon_+}\frac{\sin\left(\frac{\pi(l-t')}{L}\right)\sin\left(\frac{\pi t'}{L}\right)}{\sin\left(\frac{\pi l}{L}\right)}\right]\,,
\ee
and, in fact, we could have written previous expression \eqref{Window1} in the same way by setting $t=\sigma_1+t'$\,. Both formulas perfectly reproduce our results derived from the HRT with $h_H=\bar{h}_H$, including sub-leading corrections as for the Poincaré example. 
%%%%%%%%%%%%%%%%%%%%%%%%%%%%%
\subsubsection{CFT at finite temperature}
%%%%%%%%%%%%%%%%%%%%%%%%%%%%%
\,\,\,\,\,\,
Last but not least, we can repeat the above steps by using the maps \eqref{ExpMapsThermalCyl} to the thermal cylinder. The cross ratios become
\bea
z&=&\frac{\sinh\left(\frac{\pi(x_1+l-t+i\epsilon_-)}{\beta}\right)\sinh\left(\frac{\pi(t-x_1+i\epsilon_-)}{\beta}\right)}{\sinh\left(\frac{\pi(x_1+l-t-i\epsilon_-)}{\beta}\right)\sinh\left(\frac{\pi(t-x_1-i\epsilon_-)}{\beta}\right)}\simeq1-\frac{2\pi i\epsilon_-\sinh\left(\frac{\pi l}{\beta}\right)}{\beta\sinh\left(\frac{\pi (x_1-t)}{\beta}\right)\sinh\left(\frac{\pi (x_1+l-t)}{\beta}\right)} \,,\nonumber\\
\\
\bar{z}&=&\frac{\sinh\left(\frac{\pi(x_1+l+t-i\epsilon_+)}{\beta}\right)\sinh\left(\frac{\pi(t+x_1+i\epsilon_+)}{\beta}\right)}{\sinh\left(\frac{\pi(x_1+l+t+i\epsilon_+)}{\beta}\right)\sinh\left(\frac{\pi(t+x_1-i\epsilon_+)}{\beta}\right)}\simeq1+\frac{2\pi i\epsilon_+\sinh\left(\frac{\pi l}{\beta}\right)}{\beta\sinh\left(\frac{\pi (x_1+t)}{\beta}\right)\sinh\left(\frac{\pi (x_1+l+t)}{\beta}\right)}\,.\nn\\
\eea
After carefully examining the imaginary parts, we see that the chiral part changes the sign when $x_1<t<x_1+l$ and we can extract the form of entanglement entropy as before. We can verify that the HRT formulas are reproduced again, including sub-leading corrections. For example, the leading $\epsilon$ answer for the entangling interval on the right becomes
\be
\Delta S_A=\frac{c}{6}\log\left[\frac{\beta}{\pi\epsilon_-}\frac{\sin(\pi\alpha_-)}{\alpha_-}\frac{\sinh\left(\frac{\pi (t-x_1)}{\beta}\right)\sinh\left(\frac{\pi (x_1+l-t)}{\beta}\right)}{\sinh\left(\frac{\pi l}{\beta}\right)}\right]+O(\epsilon_-)\,,
\ee
reproducing \eqref{DSABTZRIGHT}. Analogous expression can be derived for the interval on the left, reproducing \eqref{DSABTZLEFT} in terms of $\epsilon_+$. This concludes our match with holographic 2D CFTs at large-$c$.

%%%%%%%%%%%%%%%%%%%%%%%%%%%%%%%%%%%%%%%%
%%%%%%%%%%%%%%%%%%%%%%%%%%%%%%%%%%%%%%%%
\section{Local quenches with intrinsic spin}
%%%%%%%%%%%%%%%%%%%%%%%%%%%%%%%%%%%%%%%%
%%%%%%%%%%%%%%%%%%%%%%%%%%%%%%%%%%%%%%%%
\,\,\,\,\,\,
As we have seen in the previous section, at large-$c$, 
it is straightforward to derive the evolution of entanglement entropy during local operator quenches
created by states with different conformal weights 
$(h, \bar h), h\neq \bar h$. 
This suggests that there should exist a simple holographic description of these states too. 
Since $h-\bar h = s$, these states should carry spin, so a natural description in the bulk should be in terms of infalling conical defects with spin. Indeed, these metrics were originally found in \cite{Banados:1992gq}, but to the best of our knowledge, only recently used in the $AdS/CFT$ context
\cite{Martinez:2019nor,Briceno:2024ddc,Li:2024rma,Giribet:2024nwg,Li:2026fxo}. This motivates us to reconsider these geometries in the contexts of local operator quenches here.

In this section we generalise the results of the previous
sections to infalling conical defects with spin. 
We show that the CFT description in terms of local 
quenches corresponds to left and right excitations whose heights are determined by the weights
\begin{eqnarray}
h = \frac{\Delta +s}{2} , \qquad 
\bar h = \frac{\Delta -s}{2}\,,
\end{eqnarray}
where $\Delta = mR$, with $m$ the parameter which determines the mass of the conical defect and $s$, the parameter which determines its spin. 
Interestingly, we will see that it is possible to obtain a purely left moving or a purely right moving operator quench 
by choosing extremal defects which satisfy $\Delta = |s|$. 

We will also show that the results for the entanglement entropy in large-$c$ CFTs computed using states with $h \neq \bar h$ are precisely reproduced in back-reacted 
geometries of infalling conical defects with spin. Given our detailed derivations above, we will be brief and only summarize important steps below. Once the holographic dictionary is established for a particle at the origin of $AdS_3$  with spin, then the discussion for the infalling particle with spin carries forward easily using the 
methods developed in the previous sections. 

%%%%%%%%%%%%%%%%%%%%%%%%%%%%%%%%%%%%%%%%
%%%%%%%%%%%%%%%%%%%%%%%%%%%%%%%%%%%%%%%%
\subsection{Spinning conical defect}
%%%%%%%%%%%%%%%%%%%%%%%%%%%%%%%%%%%%%%%%
%%%%%%%%%%%%%%%%%%%%%%%%%%%%%%%%%%%%%%%%
\,\,\,\,\,\,
In the previous section, we saw that the back-reaction of a massive spinless particle produces a conical defect geometry, with the strength of the defect determined by the particle's mass. If the particle also carries intrinsic spin, its back-reaction produces a spinning conical defect. This is reviewed in appendix \ref{appendix}.
The corresponding geometry is described by the metric
\begin{align}
    ds^2
    ={}& -\left(r^2+R^2-M\right)dt^2
    +\frac{R^2\,dr^2}
    {r^2+R^2-M+\dfrac{J^2}{4r^2}}
    +r^2d\phi^2-J\,dt\,d\phi .
\end{align}
The parameters $M$ and $J$ are related to the mass $m$ and intrinsic spin $s$ of the particle through
\begin{align}
    M=8G_NR^2m,
    \qquad
    J=8G_NRs.
\end{align}
We compute the boundary stress tensor using the FG expansion. First, we transform the metric into FG form, in which the radial part is separated from the tangential part. We then expand the tangential metric near the asymptotic boundary at $z=0$. The leading term determines the boundary metric, while the coefficient of the $z^2$ term determines the expectation value of the boundary stress tensor. The non-vanishing components are
\begin{align}
    T_{++}
    =\frac{M-J}{32\pi G_NR},
    \qquad
    T_{--}
    =\frac{M+J}{32\pi G_NR}.
\end{align}
In the dual CFT, this geometry corresponds to a primary state with left- and right-moving conformal weights
\begin{align}
    h=\frac{\Delta+s}{2},
    \qquad
    \bar{h}=\frac{\Delta-s}{2},
\end{align}
respectively.
%%%%%%%%%%%%%%%%%%%%%%%%%%%%%%%%%%%%%%%
\subsection*{Entanglement Entropy}
%%%%%%%%%%%%%%%%%%%%%%%%%%%%%%%%%%%%%%%
\,\,\,\,\,\,
The entanglement entropy for a spinless particle of mass $m$ located at the origin is given in \eqref{EEntropy}. When the particle carries intrinsic spin, the entanglement entropy receives an additional contribution. A convenient way to compute it is to map the spinning conical defect geometry to global AdS and then express the global AdS entanglement entropy formula in \eqref{EEntropy} in terms of the spinning-defect coordinates. The coordinate transformation we employ is
\begin{align}
    r\rightarrow\sqrt{\frac{r^2+b^2}{a^2-b^2}},\quad t\rightarrow\frac{1}{R}\lr{-a t+b\phi},\quad \phi\rightarrow\frac{1}{R}\lr{-bt+a\phi}\,.
\end{align}
Applying this transformation to the endpoints of the entangling interval and substituting the resulting expressions into \eqref{EEntropy}, we obtain
\begin{align}\label{EEsAdS}
    S_A
    =\frac{c}{6}\Bigg[
    &\log\left(\sqrt{(r_1^2+b^2)(r_2^2+b^2)}\right)
    \nonumber\\
    &+\log\left(
    \frac{
    2\cos\left(\dfrac{b\Delta\phi-a\Delta t}{R}\right)
    -2\cos\left(\dfrac{a\Delta\phi-b\Delta t}{R}\right)}
    {a^2-b^2}
    \right)
    \Bigg].
\end{align}
Here, the parameters $a$ and $b$ are determined by the conformal dimension $\Delta$ and intrinsic spin $s$ of the particle:
\begin{align}
    a
    &=\frac{R}{2}\left[
    \sqrt{1-\frac{12}{c}(\Delta-s)}
    +\sqrt{1-\frac{12}{c}(\Delta+s)}
    \right]\equiv \frac{R}{2}\left(\alpha_-+\alpha_+\right),
    \\
    b
    &=\frac{R}{2}\left[
    \sqrt{1-\frac{12}{c}(\Delta-s)}
    -\sqrt{1-\frac{12}{c}(\Delta+s)}
    \right]\equiv \frac{R}{2}\left(\alpha_--\alpha_+\right),
\end{align}
where we denoted
\be
\alpha_{\pm}=\sqrt{1-\frac{12(\Delta\pm s)}{c}}\,.
\ee
In the spinless limit, $s\to0$, we have $b\to0$, and the result reduces to the entanglement entropy of the conical defect
without spin. 
\subsection{Infalling spinning particle in the BTZ black brane}
\,\,\,\,\,\,
We will only consider quenches with spin in the background of the BTZ black brane. 
The other cases of quenches in Poincar\'{e} AdS and global AdS can be dealt with the methods developed in the previous sections.
Consider an infalling particle in the BTZ black brane with intrinsic spin $s$. The only difference with the spinless case is that the holomorphic and anti-holomorphic weight of the boundary stress tensor we get using the FG expansion are different
\bea \label{spininfallingenergy}
T_{\pm\pm}(x_\pm)=\frac{\pi R}{8G_N \beta^2}+\frac{\pi (M\mp J)}{8G_N \beta^2 R}\frac{1}{\left(\cosh\left(\frac{2\pi x_{\pm}}{\beta}\right)\cosh(\eta^\mp_{12})-\sinh(\eta^{\mp}_{12})\right)^2}\,,
\eea
This can be verified by doing a complementary CFT calculation. In the limit $\Delta=s$, the anti-holomorphic part vanishes and the stress tensor corresponding to the particle becomes chiral.
\subsection*{Entanglement Entropy for a spinning infalling Particle in BTZ}
\,\,\,\,\,\,
We obtain the entanglement entropy by applying the appropriate coordinate transformation to \eqref{EEsAdS}. An interesting feature of the spinning case is that the right- and left-moving pulses depend on the combinations $\Delta+s$ and $\Delta-s$, respectively. Consequently, in the chiral limit $\Delta=s$, only the right-moving pulse contributes to the entanglement entropy.

When both pulses lie outside the entangling interval, the entanglement entropy is
\begin{align}
   \Delta S_A
    = &\frac{\Delta+s}{3}\,\epsilon_-^2
    S_{x_1x_2}^2(t)
    +\frac{\Delta-s}{3}\,\epsilon_+^2
    S_{x_1x_2}^2(-t)\,,
\end{align}
where the time dependence of the sub-leading corrections was introduced in \eqref{epsilonsqtemp}.

On the other hand, when the right-moving pulse enters the entangling interval, we find
\begin{align}
    \Delta S_A= &\frac{c}{6}
    \log\left[
    \frac{1}{\epsilon_-}\frac{\beta}{\pi}\frac{\sin(\pi\alpha_+)}{\alpha_+}
    \frac{
    \sinh\left(\frac{\pi}{\beta}(x_1-t)\right)
    \sinh\left(\frac{\pi}{\beta}(x_2-t)\right)}
    {\sinh\left(\frac{\pi}{\beta}(x_1-x_2)\right)}
    \right].
\end{align}
Similarly, when the left-moving pulse lies inside the entangling interval, the result becomes
\begin{align}
    \Delta S_A
    = &\frac{c}{6}
    \log\left[
    \frac{1}{\epsilon_+}\frac{\beta}{\pi}\frac{\sin(\pi\alpha_-)}{\alpha_-}
    \frac{
    \sinh\left(\frac{\pi}{\beta}(x_1+t)\right)
    \sinh\left(\frac{\pi}{\beta}(x_2+t)\right)}
    {\sinh\left(\frac{\pi}{\beta}(x_1-x_2)\right)}
    \right].
\end{align}
Observe the differences in the change in the entanglement entropy in the above equations  when compared to 
the situation in which the infalling particle did not have spin, given in equations (\ref{DSABTZRIGHT}) and (\ref{DSABTZLEFT}). 
When the particle has spin, we obtain different coefficients $\alpha_+$ and $\alpha_-$ for the  change in entanglement due to the left and 
right pulses. This difference is in addition to the difference in the width of the left and right moving pulses already obtained for the spinless case. 
In fact if the spin of the infalling particle is extremal, $M =|J|$, then on re-doing the calculation, one sees there is no change in the entanglement 
for one of the intervals.  From the examination of the stress tensor in (\ref{spininfallingenergy})
we see that the pulse is either purely left or purely right moving.  

%%%%%%%%%%%%%%%%%%%%%%%%%%%%%%%%%%%%%%%%%%%%%%%%
%%%%%%%%%%%%%%%%%%%%%%%%%%%%%%%%%%%%%%%%%%%%%%%%
\section{Applications beyond holography}\label{sec:BeyondHolography}
%%%%%%%%%%%%%%%%%%%%%%%%%%%%%%%%%%%%%%%%%%%%%%%%
\,\,\,\,\,\,
So far, our discussion has focused primarily on holography and CFT states dual to particles carrying longitudinal momentum and spin in AdS. Having established the corresponding CFT dictionary, however, it is natural to explore these excited states in more general settings at the boundary. In this section, we briefly discuss three such applications: general 2D RCFTs, Free Fermion CFT, and local quantum quenches with additional momentum in generic 2D CFTs. Our discussion is intended to provide an outlook on these directions, while a more detailed analysis is left for future work.

%%%%%%%%%%%%%%%%%%%%%%%%%%%%%%%%%%%%%%%%%%%%%%%%
\subsection{Rational CFTs}
%%%%%%%%%%%%%%%%%%%%%%%%%%%%%%%%%%%%%%%%%%%%%%%%
\,\,\,\,\,\,
First, consider states excited by a local primary operator
\be
O=\sqrt{a}e^{i\sqrt{2}\tilde{\alpha}\phi}\pm\sqrt{1-a}e^{-i\sqrt{2}\tilde{\alpha}\phi}\,,
\ee
where $\phi$ is a free massless scalar. Its scaling dimensions are $h=\bar{h}=\tilde{\alpha}^2$, and parameter $a\in[0,1]$. For $a=1/2$ and $\tilde{\alpha}=1/\sqrt{8}$ the excitation corresponds to the $\sigma$ excitation in the 2D Ising CFT \cite{He:2014mwa}.

For this operator, the increase in the second Rényi entropy is given by \cite{Caputa:2015tua}
\be
\Delta S^{(2)}_A=-\log\left[a^2+(1-a)^2+2a(1-a)(|z|^{8h}+|1-z|^{8h})\right]\,.
\ee
The conformal cross-ratios $z$ and $\bar{z}$ in this formula are obtained using the map from the two-sheeted geometry to the complex plane with a branch cut along the interval $A=[u,v]$
\be
z^2=\frac{w-u}{w-v}\,.
\ee
With insertion points of local operators on the two sheets given by
\bea
&&w_2=-i(\epsilon_-+it)\,,\qquad w_1=i(\epsilon_--it)\,,\nn\\
&&\bar{w}_2=i(\epsilon_++it)\,,\qquad \bar{w}_1=-i(\epsilon_+-it)\,,\label{zsPoincare2}
\eea
on the first sheet, and points $w_3=e^{2\pi i}w_1$, $w_4=e^{2\pi i}w_2$ on the second. In this way, we get the points mapped to
\be
z_1=\sqrt{\frac{t-u+i\epsilon_-}{t-v+i\epsilon_-}}\,,\qquad z_2=\sqrt{\frac{t-u-i\epsilon_-}{t-v-i\epsilon_-}}\,,
\ee
\be
\bar{z}_1=\sqrt{\frac{t+u+i\epsilon_+}{t+v+i\epsilon_+}}\,,\qquad \bar{z}_2=\sqrt{\frac{t+u-i\epsilon_+}{t+v-i\epsilon_+}}\,,
\ee
as well as $z_3=-z_1$, $z_4=-z_2$ (and similarly for $\bar{z}_i$). 

Finally, for the interval on the right, at early or late times: $t<u$ and $t>v$, the leading $\epsilon_\pm$ expansions of the cross-ratios are
\be
z\simeq \frac{(v-u)^2\epsilon^2_-}{4(u-t)^2(v-t)^2}\,,\qquad \bar{z}\simeq\frac{(v-u)^2\epsilon^2_+}{4(u+t)^2(v+t)^2}\,.
\ee
This gives $\Delta S^{(2)}_A\simeq 0+ O(\epsilon^2_\pm)$ at these times. On the other hand, for $u<t<v$ we have
\be
z\simeq 1-\frac{(v-u)^2\epsilon^2_-}{4(t-u)^2(v-t)^2}\,,\qquad \bar{z}\simeq\frac{(v-u)^2\epsilon^2_+}{4(u+t)^2(v+t)^2}\,,
\ee
and a constant increase in the second Rényi entropy
\be
\Delta S^{(2)}_A=-\log\left[(1-a)^2+a^2\right]\,,
\ee
maximized by $a=1/2$. Note that the way this constant is approached, is controlled (to the leading order) by $\epsilon_-$. 

Similar result can be extracted when the interval is on the left $(u,v)\to (-u,-v)$, and in the non-trivial time window $|u|<t<|v|$ we get
\be
z\simeq \frac{(v-u)^2\epsilon^2_-}{4(u+t)^2(v+t)^2}\,,\qquad \bar{z}\simeq1-\frac{(v-u)^2\epsilon^2_+}{4(t-u)^2(v-t)^2}\,,
\ee
leading to the same universal constant. In this case, however, the approach to the plateau is governed by $\epsilon_+$ rather than $\epsilon_-$.

These results admit a straightforward generalization to arbitrary Rényi index $n$ by employing the monodromy properties of conformal blocks. The resulting constants are given by the logarithm of the quantum dimension of the corresponding conformal family \cite{He:2014mwa}. While the plateau value is independent of whether the interval lies to the left or right of the excitation, the regulator controlling the leading approach to this universal value differs: it is $\epsilon_-$ for intervals on the right and $\epsilon_+$ for intervals on the left.
%%%%%%%%%%%%%%%%%%%%%%%%%%%%%%%%%%%%%%%%%%%%%%%%%%%%
\subsection{Free Fermion CFT}\label{sec:free-fermion}
%%%%%%%%%%%%%%%%%%%%%%%%%%%%%%%%%%%%%%%%%%%%%%%%%%%%
\,\,\,\,\,\,
The free-fermion CFT provides another particularly simple setting in which we can study
the time evolution of R\'enyi entropies following a finite-width local quench
and compare the result with the universal prediction at order $\epsilon^2$ \cite{david2016universal}. We consider the theory on the Euclidean thermal cylinder
$\mathbb{R}\times S^1_\beta$. In Lorentzian time,
the density matrix is
\begin{equation}
    \rho(t)
    =\mathcal{N} e^{-iHt}\,
      \mathcal{O}(y_1,\bar{y}_1)\,
      \rho_\beta\,
      \mathcal{O}^{\dagger}(y_2,\bar{y}_2)\,
      e^{iHt} .
    \label{eq:free-fermion-density-matrix}
\end{equation}
Here $\mathcal{N}$ is the normalization constant. Then we consider the following quenching operator and its Hermitian conjugate in the bosonized description:
\begin{align}
    \mathcal{O}(y,\bar{y})
    &= e^{i\tilde{\alpha}[\phi(y)+\bar{\phi}(\bar{y})]}
       + k\,e^{-i\tilde{\alpha}[\phi(y)+\bar{\phi}(\bar{y})]},
    \nonumber\\
    \mathcal{O}^{\dagger}(y,\bar{y})
    &= e^{-i\tilde{\alpha}[\phi(y)+\bar{\phi}(\bar{y})]}
       + k^{*}e^{i\tilde{\alpha}[\phi(y)+\bar{\phi}(\bar{y})]} .
    \label{eq:free-fermion-operator}
\end{align}
Here, $\phi(y)$ and $\bar{\phi}(\bar{y})$ denote the holomorphic and antiholomorphic components of the bosonic field used to bosonize the fermion. We take $\tilde{\alpha}\in\mathbb{R}$, while $k$ may in general be complex. The operator has conformal weights
$(h,\bar{h})=(\tilde{\alpha}^2/2,\tilde{\alpha}^2/2)$. The state is prepared by inserting the
operator at the origin and regulating the holomorphic and antiholomorphic insertion points by $\epsilon_-$ and $\epsilon_+$, respectively.

\begin{figure}[h]
    \centering
    \includegraphics[width=0.75\textwidth]{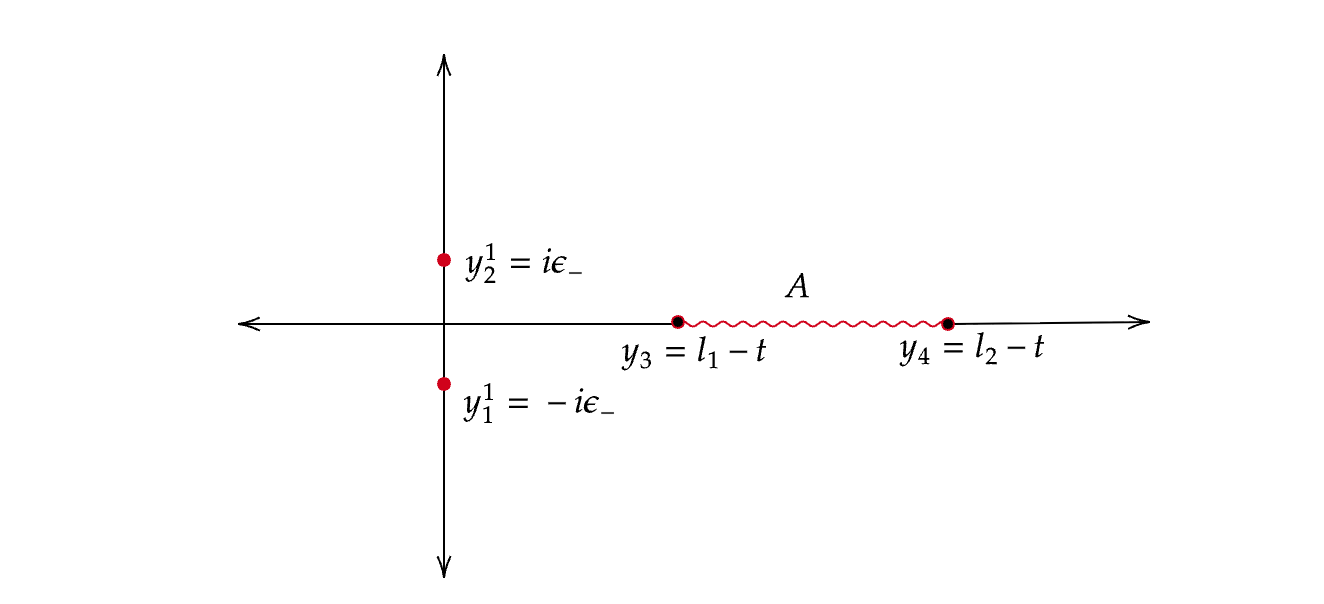}
    \caption{One sheet of the two-sheeted Riemann surface, with a branch cut
    along the interval $A=[y_3,y_4]$. The operator insertions are placed at the
    origin and regulated by $\epsilon_-$ in the holomorphic sector, with
    $y_1^{(1)}=-i\epsilon_-$ and $y_2^{(1)}=i\epsilon_-$. In the
    antiholomorphic sector, $\bar{y}_1=i\epsilon_+$ and
    $\bar{y}_2=-i\epsilon_+$.}
    \label{fig:free-fermion-riemann-sheet}
\end{figure}

We are interested in the change in the second Rényi entropy, which is obtained
from a four-point function on the two-sheeted thermal cylinder branched along
$A$:
\begin{equation}
    \Delta S_A^{(2)}
    = -\log{
        \frac{\Tr\rho_{A,\epsilon}^{2}}
             {\Tr\rho_{A}^{2}}} .
    \label{eq:second-renyi-definition}
\end{equation}
Since the four-point function is most conveniently evaluated on the complex
plane, we uniformize the branched cylinder using the map
\begin{align}
    w(y)
    &= e^{\frac{\pi}{2\beta}(y_3-y_4)}
       \left[
         \frac{\sinh\frac{\pi}{\beta}(y-y_3)}
              {\sinh\frac{\pi}{\beta}(y-y_4)}
       \right]^{\!1/2},
    \nonumber\\
    \bar{w}(\bar{y})
    &= e^{\frac{\pi}{2\beta}(\bar{y}_3-\bar{y}_4)}
       \left[
         \frac{\sinh\frac{\pi}{\beta}(\bar{y}-\bar{y}_3)}
              {\sinh\frac{\pi}{\beta}(\bar{y}-\bar{y}_4)}
       \right]^{\!1/2} .
    \label{eq:free-fermion-uniformization}
\end{align}
This map sends the branch point $y_3$ to the origin and $y_4$ to infinity.
The first and second sheets of the branched manifold are mapped to the upper and lower half-planes, respectively. Consequently, the insertion points on the second sheet are related to those on the first by a rotation of $\pi$:
\begin{equation}
    w\!\left(y_{1,2}^{(2)}\right)
       = e^{i\pi}w\!\left(y_{1,2}^{(1)}\right),
    \qquad
    \bar{w}\!\left(\bar{y}_{1,2}^{(2)}\right)
       = e^{-i\pi}\bar{w}\!\left(\bar{y}_{1,2}^{(1)}\right),
    \qquad
    w\!\left(y_p^{(1)}\right)=w_p .
    \label{eq:sheet-rotation}
\end{equation}
Using \eqref{eq:sheet-rotation}, the four-point function can be written
entirely in terms of $w_1$ and $w_2$:
\begin{align}
    &\Big\langle
       \mathcal{O}^{(1)}
       \mathcal{O}^{\dagger(1)}\mathcal{O}^{(2)}
       \mathcal{O}^{\dagger(2)}
     \Big\rangle =
    \left[
        1+|k|^4
        +2|k|^2
        \frac{|w_1+w_2|^{8\tilde{\alpha}^2}+|w_1-w_2|^{8\tilde{\alpha}^2}}
             {( |2w_1|\,|2w_2| )^{4\tilde{\alpha}^2}}
    \right] .
    \label{eq:free-fermion-four-point}
\end{align}
This expression takes a more compact form in terms of the cross-ratio
\begin{equation}
    z   = \frac{
        \sinh\frac{\pi}{\beta}(y_1-y_3)
        \sinh\frac{\pi}{\beta}(y_2-y_4)
      }{
        \sinh\frac{\pi}{\beta}(y_1-y_4)
        \sinh\frac{\pi}{\beta}(y_2-y_3)
      } .
    \label{eq:free-fermion-cross-ratio}
\end{equation}
The normalized four-point function then gives
\begin{equation}
    \Delta S_A^{(2)}
    = -\ln\!\left[
      \frac{
        1+|k|^4
        +2|k|^2\left|4\sqrt{z}\right|^{-4\tilde{\alpha}^2}
        \left(
          \left|\sqrt{z}+1\right|^{8\tilde{\alpha}^2}
          +\left|\sqrt{z}-1\right|^{8\tilde{\alpha}^2}
        \right)
      }{(1+|k|^2)^2}
    \right] .
    \label{eq:free-fermion-renyi-cross-ratio}
\end{equation}
Upon analytic continuation to Lorentzian time, $\sqrt{z}$ changes sign when
the pulse crosses an endpoint of $A$, whereas $\sqrt{\bar{z}}$ does not. In the
zero-width limit, $(\epsilon_-,\epsilon_+)\to(0,0)$, this change of branch produces a discontinuous time dependence:
\begin{equation}
    \Delta S_A^{(2;0)}
    = \begin{cases}
        0,
        & t<l_1 \ \text{or}\ t>l_2,\\[4pt]
        -\displaystyle\ln\!\left[
          \frac{1+|k|^4}{(1+|k|^2)^2}
        \right]
        +\mathcal{O}(\epsilon^4),
        & l_1<t<l_2 .
      \end{cases}
    \label{eq:free-fermion-zero-width}
\end{equation}
It is therefore natural to define the effective quantum dimension of
$\mathcal{O}$ as
\begin{equation}
    d_{\mathcal{O}}
    = \frac{(1+|k|^2)^2}{1+|k|^4},
    \label{eq:free-fermion-quantum-dimension}
\end{equation}
so that the intermediate-time plateau is
$\Delta S_A^{(2;0)}=\ln d_{\mathcal{O}}$.

At order $(\epsilon_-^2,\epsilon_+^2)$, the finite-width correction vanishes
when the pulse lies inside the interval. To display the result compactly, we
first define the outside-interval contribution
\begin{align}
    \mathcal{F}_A(t)
    &=
    \frac{\tilde{\alpha}^2\pi^2|k|^2}{\beta^2(1+|k|^2)^2}
    \sinh^2\!\left[\frac{\pi}{\beta}(l_2-l_1)\right]
    \nonumber\\[-2pt]
    &\quad\times
    \left[
      \frac{\epsilon_-^2}
      {\sinh^2\!\left[\frac{\pi}{\beta}(l_1-t)\right]
       \sinh^2\!\left[\frac{\pi}{\beta}(l_2-t)\right]}
      +
      \frac{\epsilon_+^2}
      {\sinh^2\!\left[\frac{\pi}{\beta}(l_1+t)\right]
       \sinh^2\!\left[\frac{\pi}{\beta}(l_2+t)\right]}
    \right] .
    \label{eq:free-fermion-outside-contribution}
\end{align}
The second R\'enyi entropy is then
\begin{equation}
    \Delta S_A^{(2)}
    = \begin{cases}
        \mathcal{F}_A(t)+\mathcal{O}(\epsilon^4),
        & t<l_1 \ \text{or}\ t>l_2,\\[5pt]
        -\displaystyle\ln\!\left[
          \dfrac{1+|k|^4}{(1+|k|^2)^2}
        \right]
        +\mathcal{O}(\epsilon^4),
        & l_1<t<l_2 .
      \end{cases}
    \label{eq:free-fermion-finite-width}
\end{equation}
Thus, when the pulse lies outside $A$, the correction factorizes into a universal time-dependent profile and an operator-dependent prefactor, i.e, the time dependence is universal, whereas the dependence on $k$ is not. When the
pulse lies inside the interval, the order-$\epsilon^2$ correction vanishes and
the R\'enyi entropy remains at its zero-width plateau up to
$\mathcal{O}(\epsilon^4)$. 
As mentioned before, the  time dependence of the coefficient of the $O(\epsilon^2)$ correction for pulse outside the interval $A$  is universal, and it agrees with the universal form 
in (\ref{epsilonsqtemp}) 
argued from the OPE analysis of the $2n$ point function. The prefactor, however, depends on the coefficient $k$, not just the conformal dimension of the 
operator. The reason for this is due to the fact that there is a $U(1)$ current $J$ in the fermion theory. 
Therefore, there are operators $:JJ:$ and $\partial J$ that have dimension $2$ which occur in the OPE of the primaries in (\ref{eq:free-fermion-operator}). 
The OPE has been written down in equation (3.54) of \cite{david2016universal}.
These operators gain expectation values on the replica surface and contribute along with the stress tensor. 
On evaluating these expectation values together with the stress tensor we obtain the precise coefficient of the $O(\epsilon^2)$ terms as shown in \cite{david2016universal}.
The calculation proceeds in a similar manner for quenches with different left and right moving widths. 

%%%%%%%%%%%%%%%%%%%%%%%%%%%%%%%%%%%%%%%%%%%%%%%%
\subsection{Local quenches with conserved momentum }
%%%%%%%%%%%%%%%%%%%%%%%%%%%%%%%%%%%%%%%%%%%%%%%%
\,\,\,\,\,\,
Let us now discuss another interesting application of our analysis to local quenches in 2D CFTs. In fact, a ``more universal" local quench protocol in 1+1D system is defined as follows \cite{Calabrese:2016xau}. We consider a one-dimensional system which consists of two regions $A$ and $B$, initially decoupled, with corresponding Hamiltonians $H_A$ and $H_B$. Thus, the initial state is given by
\begin{equation}
|\psi(0)\rangle=|A\rangle\otimes |B\rangle\propto \lim_{\lambda\to\infty}e^{-\lambda(H_{A}\otimes \boldsymbol{1}_B+\boldsymbol{1}_A\otimes H_{B})}|s\rangle  \,,
\end{equation}
and the choice of the state $|s\rangle$ is arbitrary, assuming $\langle s|e^{-\lambda(H_{A}\otimes \boldsymbol{1}_B+\boldsymbol{1}_A\otimes H_{B})}|s\rangle\neq 0$, such that different states only yield different normalization factors.

Then, we evolve this state with a Hamiltonian that couples $A$ and $B$. We can represent the evolution Hamiltonian as $H= H_A+ H_B+H_{AB}$, where $H_A$ and $H_B$ act non-trivially only on $A$ and $B$ respectively, while $H_{AB}$ couples the two parties. In 2D CFTs we can again use conformal maps from the plane, the cylinder or the thermal cylinder with a vertical cut (representing decoupled CFTs) to upper half plane. 

For the local quench at zero temperature we have the uniformizing functions which map the slitted geometry to the upper half plane
\be
f(w)= i\sqrt{\frac{w+i\epsilon}{i\epsilon-w}}\,,\qquad \bar{f}(\bar{w})=-i\sqrt{\frac{\bar{w}-i\epsilon}{-i\epsilon-\bar{w}}}\,.\label{MapsQPlane}
\ee
On the other hand, the thermal quench state is prepared on the cylinder with a compact Euclidean time of period $\beta$ and a slit   $\tau\in [-\beta/2,-\varepsilon]\cup[\varepsilon,\beta/2]$ at $x=0$, representing the joining of two semi-infinite halves at finite temperature at $\tau=-{\rm i}\varepsilon$. This geometry, with coordinate $w=x+{\rm i}\tau$, is then mapped to the upper half plane by
\be
\xi=f(w)={\rm i}\sqrt{\frac{\sinh\left(\frac{\pi(w+{\rm i}\varepsilon)}{\beta}\right)}{\sinh\left(\frac{\pi({\rm i}\varepsilon-w)}{\beta}\right)}}\,,\label{MapsFinTemp}
\ee
and similarly for $\bar{\xi}=\bar{f}(\bar{w})$ (see more in \cite{Caputa:2025dep}).

The expectation value of the stress tensor in CFT is consistent with the local operator quench \cite{Caputa:2014eta}, after we set the chiral dimension to $h=c/32$ ($\Delta=2h$)
\be
\langle\psi(t)| T(w)|\psi(t)\rangle=\frac{h\pi^2}{\beta^2}\frac{\sinh^2\left(\frac{2\pi {\rm i}\varepsilon}{\beta}\right)}{\sinh^2\left(\frac{\pi (w-{\rm i}\varepsilon)}{\beta}\right)\sinh^2\left(\frac{\pi (w+{\rm i}\varepsilon)}{\beta}\right)}-\frac{c\pi^2}{6\beta^2},\qquad h=\frac{c}{32}\,,
\ee
and similarly for $\bar{T}(\bar{w})$.

Finally, given our intuitions and results in this paper, a natural generalization of the local quench is simply to use maps with different chiral and anti-chiral regulators
\be
f_\pm(x_\pm)=\pm{\rm i}\sqrt{\frac{\sinh\left(\frac{\pi(x_\pm+{\rm i}\epsilon_\pm)}{\beta}\right)}{\sinh\left(\frac{\pi({\rm i}\epsilon_\pm-x_\pm)}{\beta}\right)}}\,.
\ee
The finite size result is obtained from the same maps with $\beta\to iL$, and infinite line result is recovered by taking $L\to\infty$ (or $\beta \to\infty$). A supporting evidence for this generalization is the original proposal of \cite{Nozaki:2013wia}, where infalling massive particles and their back-reacted geometries provided a qualitative description of local quenches in 2D CFTs \cite{Calabrese:2007mtj}. We expect that the conformal maps above should play a similar role for particles with conserved charges. Nevertheless, we leave interesting follow-ups, such as the derivation of entanglement entropy or generalization of this setting into spin-chains, to the upcoming works. 

Clearly, although the excited states considered here are strongly motivated by holography and by the dynamics of particles carrying conserved charges in AdS, their construction is considerably more general and can be applied to a broader class of non-equilibrium models. We have presented only three examples above, while a systematic exploration of their broader applications remains an interesting direction for future work.
%%%%%%%%%%%%%%%%%%%%%%%%%%%%%%%%%
\section{Conclusions and Outlook}\label{sec:Conclusions}
%%%%%%%%%%%%%%%%%%%%%%%%%%%%%%%%%
\,\,\,\,\,\,
In this paper we developed the holographic dictionary relating massive particles in 3 dimensional asymptotically 
AdS geometry  with conserved longitudinal or angular momenta, and spin to local quenches in CFT$_2$. 
The observation that the difference in widths of the left and right moving quenches in CFT$_2$ is related to 
conserved momenta is interesting and worth investigating further. 
Holographic quenches can be studied in higher dimensions as it was done originally in \cite{Nozaki:2013wia}. 
Here the infalling particle without momentum  is dual  to a spherical pulse of energy moving outwards from 
a given point. 
It would be interesting to study how introducing particle 
momentum changes this picture and chooses a particular direction. 
Another direction to study is  to revisit the 
study of OTOC correlators in the two-sided BTZ black hole using infalling particles 
with momenta as was done earlier for quenches carrying only energy  in \cite{Caputa:2015waa}. 

More broadly, understanding localized excitations carrying conserved charges is expected to play an important role in holography and quantum gravity. Such configurations provide elementary building blocks for more general dynamical processes, including scattering in AdS, black-hole formation and evaporation, and the transport of quantum information in gravitational systems. We hope that the framework developed here will provide a useful starting point for future investigations of non-equilibrium dynamics, quantum chaos, scrambling and the role of conserved charges in the emergence of semiclassical spacetimes. 

%%%%%%%%%%%%%%%%%%%%%%%%%%%%%
%%%%%%%%%%%%%%%%%%%%%%%%%%%%%
\acknowledgments
\,\,\,\,\,\,
We would like to thank Sumit R. Das, Giuseppe Di Giulio, Joan Simon, Tadashi Takayanagi for discussions and Giuseppe Di Giulio and Masahiro Nozaki for comments on the draft.
PC and PCG are supported by the  Swedish Research Council (VR) under Grant No. 2025-04154 and the ERC Consolidator grant (number: 101125449/acronym: QComplexity). Views and opinions expressed are however those of the authors only and do not necessarily reflect those of the European Union or the European Research Council. Neither the European Union nor the granting authority can be held responsible for them. JRD is partially supported by ANRF, India: grant no.  ANRF/ARGM/2025/00054/TS. RM acknowledges
support of the Department of Atomic Energy, Government of India, under project no.
RTI4019.

\appendix 
%%%%%%%%%%%%%%%%%%%%%%%%%%%%%%%%%%%%%%%%
\section{Back-reacted metric for spinless and spinning static particle} \label{appendix}
%%%%%%%%%%%%%%%%%%%%%%%%%%%%%%%%%%%%%%%%
\,\,\,\,\,\,
We first review the derivation of the metric of the massive particle at the origin of global $AdS_3$ given in (\ref{BRMetric}). This was originally done in 
\cite{Deser:1983nh}.
The stress tensor of the massive particle  following the trajectory $\tilde x^\mu (\lambda) $ is given by \cite{Weinberg:1972kfs} (page 44):
\begin{align}
    T^{\mu\nu} (x) = m\int d\lambda\ \frac{\dot{\tilde{x}}^\mu\ \dot{\tilde{x}}^\nu}{e(\lambda)} \ \frac{\delta^3\left( x-\tilde{x}(\lambda)\right)}{\sqrt{g}}, 
    %\quad u^\mu=\frac{\dot{\tilde{x}}^\mu}{e(\lambda)}, 
    \qquad e(\lambda)=\sqrt{-g_{\mu\nu}\dot{\tilde{x}}^\mu\dot{\tilde{x}}^\nu}.
\end{align}
Here $\lambda$ parametrizes the world line  of the particle. 
The world line of a particle without spin at the origin of 
 $AdS_3$ is given by 
\begin{align} \label{worldline}
    \tilde{t}=m\lambda,\quad \tilde{r}=0,\quad \tilde{\phi}=0.
\end{align}
Let us assume that the back-reacted metric is of the form 
\begin{align}\label{ansatz}
    ds^2 = -\lr{r^2 + R^2 + G_N f_1(r)} dt^2 + \frac{ R^2 dr^2}{\lr{r^2 + R^2 + G_N f_2(r)}} + r^2 d\phi^2,
\end{align}
To begin with, we will solve 
 Einstein's equations to the linear order in $G_N$. 
Substituting the ansatz (\ref{ansatz}) in the equation for the stress tensor,  we see that the  only non-vanishing component of the stress tensor is  the $(tt)$ component which is given by
\begin{eqnarray}\label{Ttt}
    T^{tt} &=& m\int d\lambda\ \dot{\tilde{t}}^{~2}\  \frac{\delta\left(t-\tilde{t}(\lambda)\right)\delta( r)\delta(\phi) \ }{m  R^2 r} , \\
    \nonumber
    && 
    = \frac{m}{ R^2 }\frac{\delta(r)\delta(\phi)}{r}.
\end{eqnarray}
Here we have used the fact that we need the stress tensor 
to the zeroth order in $G_N$ to solve the Einstein's equations which are given by 
\begin{align}
    R_{\mu\nu} - \frac{1}{2} g_{\mu\nu} R - \frac{1}{R^2}g_{\mu\nu} = 8\pi G_{\text{\tiny N}} T_{\mu\nu}.
\end{align}
\noindent

The components of the Einstein tensor to order $G_N$ with the cosmological constant are given by 
\begin{align}
    G_{tt}=-\frac{R^2+r^2}{2rR^2}f_2'(r)G_N, \quad G_{rr}=\frac{ \left(r^2+R^2\right) f'_1(r)-2 r f_1(r)+2 r f_2(r)}{2 r \left(r^2+R^2\right)^2}G_N\nonumber\\
   G_{\phi\phi}= \frac{r^2 \left(\left(r^2+R^2\right) \left(\left(r^2+R^2\right) f_1''(r)+r \left(f_2'(r)-f_1'(r)\right)\right)-2 R^2 f_1(r)+2 R^2 f_2(r)\right)}{2 R^2
   \left(r^2+R^2\right)^2}G_N.
\end{align}
Therefore, the components of Einstein's equation to the 
leading order in $G_N$ are the following
\begin{align}
    G_{tt}=8\pi G_N T_{tt},\quad G_{rr}=8\pi G_N T_{rr},\quad G_{\phi\phi}=8\pi G_N T_{\phi\phi},\quad T_{\mu\nu}=g_{\mu\sigma}g_{\nu\rho}T^{\sigma\rho}.
\end{align}
As the equations are spherically symmetric, we first integrate over $\phi$ and then solve the differential equation. We get the $f_2$ by solving for $G_{tt}$
\begin{align}
     f_2(r>0) = -8 m \int_0^r dr\ \delta(r)(R^2 + r^2)= -8 mR^2.
     \end{align}
It is clear from this equation that $f_2(r)$ is discontinuous at the origin. 
\noindent
Substituting the value of $f_2(r)$ in $G_{rr}$, we obtain a ordinary differential equation
\begin{align}
   % \frac{-2 r f_1(r) + 2 r f_2(r) + (R^2 + r^2) f_1'(r)}{2r %(R^2 + r^2)^2} = 0 \implies 
    (R^2 + r^2) f_1'(r) - 2r f_1(r) = 16 m r R^2. 
\end{align}
We solve this differential equation assuming asymptotic AdS boundary conditions 
\begin{eqnarray}
 f_1(r) &=& (R^2 + r^2) \int_0^r dr \frac{16 m rR^2 }{( R^2 + r^2)^2} , \\ \nonumber
 &=& 8 m R^2(R^2 + r^2) \lrt{-\frac{1}{R^2 + r^2}}=-8m R^2  . 
\end{eqnarray}
Observe  that $f_1(r) = f_2(r), r>0$ and that the $(\phi, \phi)$ component of the Einstein equation is satisfied.
Therefore the point particle   deforms global $AdS_3$  into a conical deficit geometry which is given by 
\begin{align}\label{backreacted metric of AdS}
    ds^2= -\lr{r^2 + R^2 - \tilde{m}} dt^2 + \frac{R^2dr^2}{r^2 + R^2 - \tilde{m}} + r^2 d\phi^2, \quad \tilde{m}=8G_NR^2m.
\end{align}
Although  this metric was obtained using an analysis to the leading order in $G_N$, it can be shown that 
the condition  $f_1(r) = f_2(r) = - 8 m R^2, r> 0$, 
ensures that the metric in 
(\ref{backreacted metric of AdS}) is  exact solution to the Einstein equations of motion 
with the point source \cite{Deser:1983nh}. 

\subsubsection*{Spinning Particle}
\,\,\,\,\,\,
For a spinning particle, we expect  the back-reacted metric
to be that of the spinning conical defect  found 
in \cite{Banados:1992gq}.
In this  section of the appendix, we show
that this is indeed the case. 
A spinning particle  contains an additional term
in its stress tensor proportional to the spin
\cite{Faye:2006gx,Raeymaekers_2023}.

Let the  world line  of the particle   be $\tilde{x}^{\mu}(\lambda)$, where
$\lambda$ is an arbitrary, not necessarily  the  affine parameter. Define
\begin{align}
    \dot{\tilde{x}}^{\mu}
    &\equiv \frac{d\tilde{x}^{\mu}}{d\lambda},
    &
    e(\lambda)
    &\equiv \frac{d s}{d\lambda}
    =\sqrt{
        -g_{\mu\nu}\left(\tilde{x}(\lambda)\right)
        \dot{\tilde{x}}^{\mu}
        \dot{\tilde{x}}^{\nu}
    },
\end{align}
Here $u^\mu$ is the proper velocity  
\begin{align} \label{propervel}
    u^{\mu}
    =\frac{d\tilde{x}^{\mu}}{d s}
    =\frac{\dot{\tilde{x}}^{\mu}}{e(\lambda)},
    \qquad
    u^{\mu}u_{\mu}=-1.
\end{align}
The stress tensor along with the contribution from the spin 
is given by 
\begin{align}
\begin{aligned}
    T^{\mu\nu}(x)
    =&
    m\int d\lambda\,
    \frac{
        \dot{\tilde{x}}^{\mu}(\lambda)
        \dot{\tilde{x}}^{\nu}(\lambda)
    }{e(\lambda)}
    \frac{
        \delta^{(3)}\!\left(
            x-\tilde{x}(\lambda)
        \right)
    }{\sqrt{-g}}
    \\
    &-\nabla_{\rho}\int d\lambda\,
    \frac{1}{2}
    \left(
        S^{\rho\mu}\dot{\tilde{x}}^{\nu}
        +
        S^{\rho\nu}\dot{\tilde{x}}^{\mu}
    \right).
    \frac{
        \delta^{(3)}\!\left(
            x-\tilde{x}(\lambda)
        \right)
    }{\sqrt{-g}} ,
\end{aligned}
\end{align}
here $S^{\mu\nu}$ is the antisymmetric spin tensor. In three dimensions it is given by 
\begin{align}
S^{\mu\nu}=s\epsilon^{\mu\nu\sigma}u_{\sigma}
,\quad 
\epsilon^{\mu\nu\sigma}=\frac{\tilde{\epsilon}^{\mu\nu\sigma}}{\sqrt{-g}}, \qquad \tilde\epsilon^{r\phi t } =1.
\end{align}

We begin with the following ansatz for the metric which is 
a modified from that in (\ref{ansatz})
\begin{align}\label{ansatz2}
    ds^2 = -&\lr{r^2 + R^2 + G_N f_1(r)} dt^2 + R^2\frac{dr^2}{\lr{r^2 + R^2 + f_2(r) G_N+g_1(r) G_N^2}} \nonumber\\
    &+ r^2 d\phi^2 +  g(r)G_N\ dt d\phi.
\end{align}
Note that we expect a  $(t\,, \phi)$ component due to the contribution of the stress tensor from the spin in these directions. Let us evaluate the stress tensor for the particle following 
the world line in (\ref{worldline}). 
Since it is sufficient to solve the 
Einstein's equations to the leading 
order in $G_N$, the stress tensor can be evaluated in the background metric (\ref{ansatz2}) ignoring all terms involving  $G_N$. 
Using the definition of the proper velocity in (\ref{propervel}), 
we see that 
\begin{eqnarray}
    u_\mu = ( - R, 0, 0 ),
\end{eqnarray}
and the non-zero components of  spin tensor from (\ref{defispin}) reduces to  
\begin{align} \label{defispin}
   S^{r\phi}=s\frac{\epsilon^{r\phi t}}{Rr}u_t=-\frac{ s}{r},\quad S^{\phi r}=\frac{ s}{r}.
\end{align}
The first part of the stress tensor comes from the mass of the particle and contributes to the component $T^{tt}$, we have evaluated this earlier in (\ref{Ttt}). 
To obtain the contribution of the stress tensor from the
spin we 
compute
\begin{align}
    \nabla_r
\left(\frac{1}{2}
    \left(
        S^{\rho\mu}\dot{\tilde{x}}^{\nu}
        +
        S^{\rho\nu}\dot{\tilde{x}}^{\mu}
    \right).
\frac{\delta^{(3)}\!\left(x-\tilde{x}(\lambda)\right)}{\sqrt{-g}}
\right)&=\nabla_r\lr{
\frac{S^{r\phi}\dot{\tilde{x}}^t}{2} \frac{\delta^{(3)}\!\left(x-\tilde{x}(\lambda)\right)}{\sqrt{-g}}}\nonumber\\
&=-\frac{ms}{2r}\delta(t-m\lambda)\partial_r\lr{\frac{\delta(r)\delta(\phi)}{Rr}}.
\end{align}
We have used the fact that 
\begin{align}
\nabla_r\lr{{S^{r\phi}\dot{\tilde{x}}^t}}=\partial_r\lr{{S^{r\phi}\dot{\tilde{x}}^t}}+ \Gamma_{r\alpha}^r{S^{\alpha\phi}\dot{\tilde{x}}^t}+ \Gamma_{r\alpha}^\phi{S^{r\alpha}\dot{\tilde{x}}^t}+\Gamma_{r\alpha}^t{S^{r\phi}\dot{\tilde{x}}^\alpha} =0.
\end{align}
Therefore  the non-zero components are
\begin{align}\label{spinning_ST}
    T^{tt}=\frac{m}{R^2}\frac{\delta(r)\delta(\phi)}{r},\quad T^{t\phi}=\frac{s}{2r}\partial_{r}\lrt{\frac{\delta(r)\delta(\phi)}{Rr}}.
\end{align}

We proceed to evaluate the  Einstein tensor and equate it to the stress tensor given in \eqref{spinning_ST}. Non-zero components are
\begin{align}
    G_{tt}&=-\frac{G_N \left(r^2+R^2\right) f_2'(r)}{2 r R^2},\quad G_{t\phi}=-\frac{G_N \left(r^2+R^2\right) \left(r g''(r)-g'(r)\right)}{4 r R^2},\nonumber\\
    G_{\phi\phi}&=\frac{r^2 G_N \left(\left(r^2+R^2\right) \left(\left(r^2+R^2\right)
   f_1''(r)+r \left(f_2'(r)-f_1'(r)\right)\right)-2 R^2 f_1(r)+2 R^2
   f_2(r)\right)}{2 R^2 \left(r^2+R^2\right)^2},\\
  G_{rr}&=\frac{
   \left(r^2+R^2\right) f_1'(r)-2 r f_1(r)+2 r f_2(r)}{2 r
   \left(r^2+R^2\right)^2}G_N\nonumber\\ 
   &+ \frac{-8 r f_1(r) \left(r^2+R^2\right) f_1'(r)+16 r^2
   f_1(r){}^2-16 r^2 f_2(r){}^2}{16 r^2 \left(r^2+R^2\right)^3}G_N^2 \\ \nonumber
   &+\frac{\left(r^2+R^2\right)
   \left(\left(r^2+R^2\right) g'(r)^2+16 r^2 g_1(r)-4
   g(r)^2\right)}{16 r^2 \lr{r^2+R^2}^3} G_N^2.
\end{align}
First we solve for the $G_{t\phi}$ component
\begin{align}
    -\frac{G_N \left(r^2+R^2\right) \left(r g''(r)-g'(r)\right)}{4 r R^2}=-8\pi G_N\times r^2\lr{r^2+R^2} \lr{\frac{s}{2r}} \partial_r\lr{\frac{\delta(r)\delta(\phi)}{Rr}}.
\end{align}
Integrating over $\phi$ and rearranging terms we write 
\begin{align}
    \int_0^r dr \lr{rg'' - g'}=8sR\int dr\lrt{r^2\partial_r\lr{\frac{\delta(r)}{r}}}\implies rg'-2g=16sR.
    \end{align}
    Solving the differential equation using asymptotically 
AdS boundary conditions we find
    \begin{align} \label{solutiong}
        g(r)=r^2\int_0^r\frac{16sR}{r^3}dr=-8Rs .
    \end{align}
From the $(tt)$ component of the Einstein equation 
we can solve for the 
 function $f_2(r)$, which is identical to the 
 case without spin. The $(rr)$ component of the 
 Einstein equation can be used to solve for $g_1(r)$. This results in 
\begin{align} \label{solutionf1f2g1}
    f_1(r)=f_2(r) = -8 m R^2 ,\quad  \text{and}\quad  g_1(r)= \frac{g(r)^2}{4r^2}.
\end{align}
Note that the $(\phi\phi)$ component of the Einstein equations is satisfied with these solutions. 
Although we have solved the Einstein's equation to the leading order in $G_N$, it can be shown that the solutions for the unknown functions in (\ref{solutiong}), (\ref{solutionf1f2g1}) satisfy the Einstein's equations identically to all orders in $G_N$ for $r>0$. 
Therefore the back-reacted metric for the particle 
carrying intrinsic spin is given by 
\begin{align}
    ds^2
    ={}& -\left(r^2+R^2-\widetilde{m}\right)dt^2
    +\frac{R^2\,dr^2}
    {r^2+R^2-\widetilde{m}+\dfrac{J^2}{4r^2}}
    +r^2d\phi^2-J\,dt\,d\phi .
\end{align}
The parameters $\widetilde{m}$ and $J$ are related to the mass $m$ and intrinsic spin $s$ of the particle through
\begin{align}
    \widetilde{m}=8G_NR^2m,
    \qquad
    J=8G_NRs.
\end{align}

%%%%%%%%%%%%%%%%%%%%%%%%%%%%%%%%%
\appendix
%%%%%%%%%%%%%%%%%%%%%%%%%%%%%
%%%%%%%%%%%%%%%%%%%%%%%%%%%%
\bibliography{RefsK}
\bibliographystyle{JHEP}

\end{document}